%% file: main.tex
\documentclass[journal]{IEEEtran}

\usepackage{Packages}
\usepackage{titlesec}
\usepackage{indentfirst}
\titlespacing*{\section}{2pt}{2pt}{2pt}
\titlespacing*{\subsection}{2pt}{2pt}{2pt}
\titlespacing*{\subsubsection}{0pt}{2pt}{1pt}

\AtBeginDocument{
    \setlength{\abovedisplayskip}{2pt}
    \setlength{\belowdisplayskip}{2pt}
    \setlength{\abovedisplayshortskip}{2pt}
    \setlength{\belowdisplayshortskip}{2pt}
}

\usepackage{graphicx}
\usepackage{capt-of}

\def\BibTeX{{\rm B\kern-.05em{\sc i\kern-.025em b}\kern-.08em
    T\kern-.1667em\lower.7ex\hbox{E}\kern-.125emX}}

\begin{document}
\include{Definition}
\providecommand{\blue}[1]{#1}
\providecommand{\red}[1]{#1}
\renewcommand{\blue}[1]{#1}
\renewcommand{\red}[1]{#1}

\colorlet{blue}{black}
\colorlet{red}{black}
\colorlet{purple}{black}
\title{Energy Efficiency Maximization for Hybrid RIS-Aided Communications via Deep Unfolding\vspace{-5pt}}

\author{Pouya~Mobaraki, {\it Graduate Student Member, IEEE}, Abolfazl~Zakeri, {\it Member, IEEE}, Marco~Di~Renzo, {\it Fellow, IEEE}, Markku~Juntti, {\it Fellow, IEEE}, and Nhan~Thanh~Nguyen, {\it Senior Member, IEEE}%
\thanks{
Pouya Mobaraki, Abolfazl Zakeri, Markku Juntti, and Nhan Thanh Nguyen are with the Centre for Wireless Communications, University of Oulu, P.O. Box 4500, FI-90014 Oulu, Finland (e-mail: \{pouya.mobaraki, abolfazl.zakeri, markku.juntti, nhan.nguyen\}@oulu.fi). 
Marco Di Renzo is with CNRS, CentraleSup\'elec, Laboratoire des Signaux et Syst\`emes, Universit\'e Paris-Saclay, 91192 Gif-sur-Yvette, France, and also with the Centre for Telecommunications Research, Department of Engineering, King's College London, WC2R 2LS London, U.K. (e-mail: marco.1.di\_renzo@kcl.ac.uk).
\\ Preliminary results of this work were presented in \cite{Pouya_conf}.
}
}
\markboth{}%
{How to Use the IEEEtran \LaTeX \ Templates}
\IEEEaftertitletext{\vspace{-20pt}}
\maketitle
\begin{abstract} We address energy-efficiency (EE) maximization in a multiuser (MU) multiple-input single-output (MISO) downlink system assisted by a hybrid reconfigurable intelligent surface (RIS), where the positions of a predefined number of active elements are optimized while the remaining elements operate in passive mode. Practical hardware effects are explicitly incorporated, including base station (BS) and RIS power budgets, active-element amplitude-coefficient limits, amplification noise, and binary phase control. To solve the problem, we develop an alternating-optimization framework in which the BS beamforming subproblem is handled via zero-forcing with closed-form power allocation, while the RIS subproblem is addressed using a model-driven deep unfolding approach. Numerical results show that the proposed method achieves faster convergence and higher EE than the considered benchmark schemes. In particular, it attains about $20 \%$ higher EE and is five times faster than the procedure without deep unfolding. Furthermore, our simulations demonstrate at least a $5 \%$ EE improvement over the fully active RIS configuration with lower amplifier hardware requirements, and up to fivefold EE gains compared with the fully passive RIS design. The results also show that most of the achievable EE gain can be captured by activating only a small fraction of the RIS elements and allocating only a small portion of the dynamic power budget to the RIS.
\end{abstract}
\section{Introduction}
Reconfigurable intelligent surfaces (RISs) have attracted significant interest as a promising technology for future wireless systems due to their ability to modulate the wireless channel in a programmable manner~\cite{wymeersch2020radio,umer2025reconfigurable,9140329}. By adjusting the phase of incident electromagnetic waves, RISs can redirect reflected signals toward intended users and improve coverage, spectral efficiency (SE), and energy efficiency (EE)~\cite{huang2019reconfigurable}. However, when the incident signal is weak, conventional ``passive'' RIS cannot compensate for the severe multiplicative attenuation introduced by the cascaded transmitter--RIS--receiver channels. As a result, the received signal strength may become insufficient to guarantee reliable communications or the required quality of service (QoS)~\cite{zhang2022active}. To overcome this limitation of passive RIS designs, ``active'' RIS architectures were introduced, in which reflection-type amplifiers are integrated into the RIS to amplify the incoming signal before reflection, thereby alleviating cascaded channel~attenuation~\cite{liu2021active}.
\\\indent
However, equipping every RIS element with an amplifier significantly increases circuit power consumption, hardware complexity, and amplification noise. To achieve a balance among communication performance, power consumption, and hardware complexity, Nguyen \textit{et al.}~\cite{nguyen2022hybrid} 
introduced a ``hybrid'' RIS architecture, where only a subset of elements operate in active mode, while the remaining elements act as passive reflectors. Nevertheless, hybrid RIS design, particularly under the EE criterion, requires careful selection of the active RIS elements, since each active element needs a dedicated amplifier and additional circuit power. As a result, activating more elements increases total power consumption, while the corresponding improvement in communication performance (i.e., SE) may become negligible. 
This, in turn, leads to a noticeable reduction in the system EE.
\\\indent 
EE optimization for hybrid RIS-assisted systems has recently been studied in~\cite{rodrigues2023efficient,lyu2023energy,huang2019reconfigurable,choudhury2026energy,huang2024hybrid,le2024distributed,xie2025energy,peng2023hybrid}. These works mainly employ conventional iterative methods, such as alternating optimization (AO), block coordinate descent, and successive convex approximation (SCA), and demonstrate the EE gains achievable with hybrid RIS. 
Nevertheless, several of these works assume that the active/passive operating mode of each RIS element is predetermined, rather than optimized as part of the system design. This assignment directly shapes the cascaded channel and, consequently, the achievable EE. However, jointly optimizing the positions of a fixed number of active elements with the continuous beamforming and RIS coefficients renders the problem computationally demanding and leads to slow convergence under conventional iterative methods.
Motivated by this, we address the EE maximization problem for hybrid RIS-assisted communications, where the positions of a fixed number of active RIS elements are optimized while the remaining elements operate in passive mode, using a \textit{model-driven deep unfolding}~approach~\cite{hershey2014deep,shlezinger2025deep}.

We consider a hybrid RIS-assisted multiuser (MU) multiple-input single-output (MISO) downlink system, as illustrated in Fig.~\ref{fig:system_model}. Our objective is to jointly design the transmit beamforming at the BS, the positions and amplitude coefficients of a predefined number of active RIS elements, and the phase shifts of all RIS elements to maximize the system EE. The design is subject to the BS and RIS power budgets, per-user minimum-SE QoS constraints, the BS and RIS power consumption models, and \textit{binary} phase-shift constraints of the RIS~elements.

The resulting problem is a mixed-integer non-convex optimization problem with a fractional EE objective, coupled variables, non-convex constraints, and discrete active-element set selection. To address this, we propose two methods. Our primary method employs an AO framework to decouple the problem into two parts: BS beamforming and RIS optimization. The BS beamforming subproblem is addressed using a zero-forcing (ZF) beamformer, with power allocation obtained in closed form using the Dinkelbach technique. For the RIS optimization part, we propose a model-driven deep unfolding approach in which the iterations of the projected gradient method are unrolled into trainable layers, with smoothing and relaxation techniques introduced to enable differentiable treatment of the discrete active-element selection and binary phase-shift constraints. As a benchmark, the second method applies conventional projected gradient ascent (PGA), where all optimization variables are jointly updated through gradient ascent iterations with projection onto the feasible sets; the QoS constraints are enforced via a barrier penalty incorporated into~the~objective.

We conduct extensive simulations to evaluate the proposed deep unfolding-based AO method and the PGA benchmark, alongside different RIS architectures, including fully passive and fully active RISs. We further analyze the impact of key system parameters on EE, including the number of active elements and the power budget allocation between the BS~and~the~RIS.

\blue{Our main contributions are summarized as follows:
\begin{itemize}
\item We address an EE maximization problem for a hybrid RIS-assisted MU-MISO downlink system by jointly optimizing the positions of a fixed number of active RIS elements, BS beamforming, and RIS coefficients under power and quality-of-service (QoS)~constraints.
\item We develop a model-driven deep unfolding approach for RIS optimization, combining learnable PGD step sizes with smoothing and relaxation techniques to handle discrete active-element selection and binary phase-shift~constraints.
\item We further develop a PGA-based benchmark that jointly optimizes BS beamforming and RIS coefficients using derived gradients, with active-element selection handled through projection and the QoS constraint enforced using a barrier~penalty method.
\item We present numerical results demonstrating that the proposed deep unfolding-based AO method achieves higher EE and faster convergence than fixed-step and backtracking AO-PGD, and nearly twofold higher EE than the PGA-based method. Furthermore, the hybrid RIS outperforms fully passive and fully active architectures, while activating only 20\% of the elements achieves approximately 80\% of the maximum EE under the considered~settings.
\end{itemize}}

\textbf{Organization:}
The rest of this paper is organized as follows. Section~\ref{sec:relat} reviews the related work. Section~\ref{sec:model} presents the system model and problem formulation. Section~\ref{sec:deepAO} presents the proposed deep unfolded AO method to solve the resulting problem. Section~\ref{sec:grad} presents the PGA-based solution to the main problem. Simulation results are provided in Section~\ref{sec:result}, followed by the conclusions in Section~\ref{sec:conc}.

\textbf{Notations:} Scalars, vectors, matrices, and sets are denoted by lower-case letters, bold-face lower-case letters, bold-face upper-case letters, and calligraphic upper-case letters, respectively. The conjugate and conjugate transpose are respectively denoted by $(\cdot)^\ast$ and $(\cdot)^\H$. Additionally, $\mathbb{C}^{m\times n}$ denotes the set of $m\times n$ complex-valued matrices. $\mI_N$ denotes the $N\times N$ identity matrix, and $\diag{a_1,\ldots,a_N}$ denotes a diagonal matrix with diagonal entries $a_1,\ldots,a_N$. The operators $\tr{\cdot}$ and $\nabla$ denote the trace and gradient, respectively. Moreover, $|\cdot|$ denotes the absolute value of a scalar, $\|\cdot\|$ denotes the Euclidean norm, and $\|\cdot\|_F$ denotes the Frobenius norm. The cardinality of a set $\mathcal{B}$ is denoted by $|\mathcal{B}|$, and $\mathbbm{1}_N^{\mathcal{A}}$ denotes the $N$-dimensional indicator vector associated with the set $\mathcal{A}$. Finally,
$\mathcal{CN}(\mathbf{0},\mC)$ denotes the circularly symmetric complex Gaussian distribution with zero mean and covariance matrix $\mC$.


\section{Related Work} \label{sec:relat}
Hybrid RIS-assisted transmission design has been extensively studied under different system configurations and performance metrics~\cite{nguyen2022hybrid,ju2024beamforming,kang2023active,nguyen2024fairness,xie2024exploring,nguyen2022hybridactive}. In~\cite{nguyen2022hybrid}, a hybrid RIS-assisted multiple-input multiple-output (MIMO) system was considered using an AO framework with SCA, showing that a small number of active elements can significantly improve SE compared with fully passive RIS designs. Similarly,~\cite{ju2024beamforming} investigated weighted sum-rate maximization in an MU-MIMO system using an SCA-based AO approach, demonstrating that hybrid RIS can effectively balance passive beamforming gains and active signal amplification while mitigating the multiplicative path-loss effect. The work in~\cite{kang2023active} analyzed the max-min ergodic capacity of a hybrid RIS-assisted multiuser system by optimizing RIS phase shifts, amplification coefficients, and active-element positions. In~\cite{nguyen2024fairness}, a hybrid RIS-assisted unmanned aerial vehicle (UAV) communication system was studied, where BCA and SCA were employed to maximize the minimum user rate through joint optimization of the UAV trajectory and RIS~coefficients.

Beyond SE, EE in RIS-assisted systems has also received considerable attention~\cite{huang2019reconfigurable,lyu2023energy,choudhury2026energy,huang2024hybrid,le2024distributed,xie2025energy,peng2023hybrid,soleymani2025spectral,fotock2023energy}. In~\cite{huang2019reconfigurable}, EE was maximized for a fully passive RIS-assisted MU-MISO downlink using fractional programming within an AO framework. For hybrid RIS systems,~\cite{lyu2023energy} investigated a cell-free network and demonstrated that a small number of active elements can outperform fully passive and fully active architectures. Other studies considered EE optimization in massive MIMO systems with hardware impairments~\cite{choudhury2026energy}, hybrid RIS transmitters~\cite{huang2024hybrid}, dynamic mode switching~\cite{xie2025energy}, and active/passive element allocation~\cite{peng2023hybrid}. These works demonstrate the EE benefits of hybrid RIS, mainly through conventional iterative optimization methods.

One important aspect of hybrid RIS design is the incorporation of practical hardware constraints, particularly low-resolution phase control. In this regard, one-bit phase shifts have been recognized as a cost-effective and energy-efficient solution~\cite{stylianopoulos2022online}. For example,~\cite{sang2023quantized} investigated discrete phase-shift design and proposed efficient quantization methods. Beyond phase control, active/passive element assignment also significantly affects system performance. In~\cite{zappone2020optimal}, the rate, EE, and their trade-off were optimized with respect to the number and phase shifts of RIS elements while accounting for channel estimation and configuration overheads. In~\cite{gong2024hybrid}, fixed and dynamic hybrid RIS architectures were studied for secure communications by jointly optimizing the BS transmit covariance, RIS coefficients, and element allocation. Similarly,~\cite{xie2024exploring} investigated flexible mode switching in mobile edge computing systems and demonstrated its impact on the latency-energy trade-off.

Nevertheless, jointly optimizing binary phase shifts and active/passive element selection introduces significant computational challenges, particularly in large-scale RIS systems. To address such challenges, data-driven deep learning approaches have been proposed for RIS optimization~\cite{zhong2021ai}. However, these methods often suffer from limited interpretability and dependence on large training datasets~\cite{wu2023deep}. In contrast, model-based deep learning, particularly deep unfolding, incorporates the iterations of conventional optimization algorithms into trainable neural network layers~\cite{hershey2014deep,balatsoukas2019deep,shlezinger2025deep}. This approach preserves the underlying optimization structure while improving convergence through learned parameters. Deep unfolding has also been applied to RIS-assisted systems for transmit beamforming and joint active/passive coefficient optimization~\cite{zhang2024joint,nguyen2024joint}. However, these studies do not jointly address discrete active-element selection and binary phase-shift constraints within an EE optimization framework. These coupled discrete constraints introduce non-differentiability, motivating the development of a tailored deep-unfolding approach for the considered problem.

\blue{The most closely related works to ours are~\cite{choudhury2026energy,huang2024hybrid,xie2025energy,peng2023hybrid}. In~\cite{choudhury2026energy}, EE optimization with discrete RIS phase shifts and hardware impairments were investigated without optimizing active-element locations. Huang \textit{et al.}~\cite{huang2024hybrid} considered EE optimization for a hybrid RIS transmitter using AO-based methods, while~\cite{xie2025energy} investigated dynamic mode switching in hybrid RIS using conventional iterative optimization. In~\cite{peng2023hybrid}, the number of active and passive RIS elements was optimized to maximize EE. In contrast, our work jointly considers active-element location selection, binary phase shifts, and continuous amplification and beamforming variables within an EE optimization framework. More importantly, we develop a tailored deep-unfolding-based AO approach to handle the resulting mixed discrete--continuous problem.}
\section{System Model and Problem Formulation} \label{sec:model}
\subsection{System Model}
\begin{figure}[t]
    \vspace{-0.3cm}
    \centering
    \includegraphics[width=0.95\linewidth]{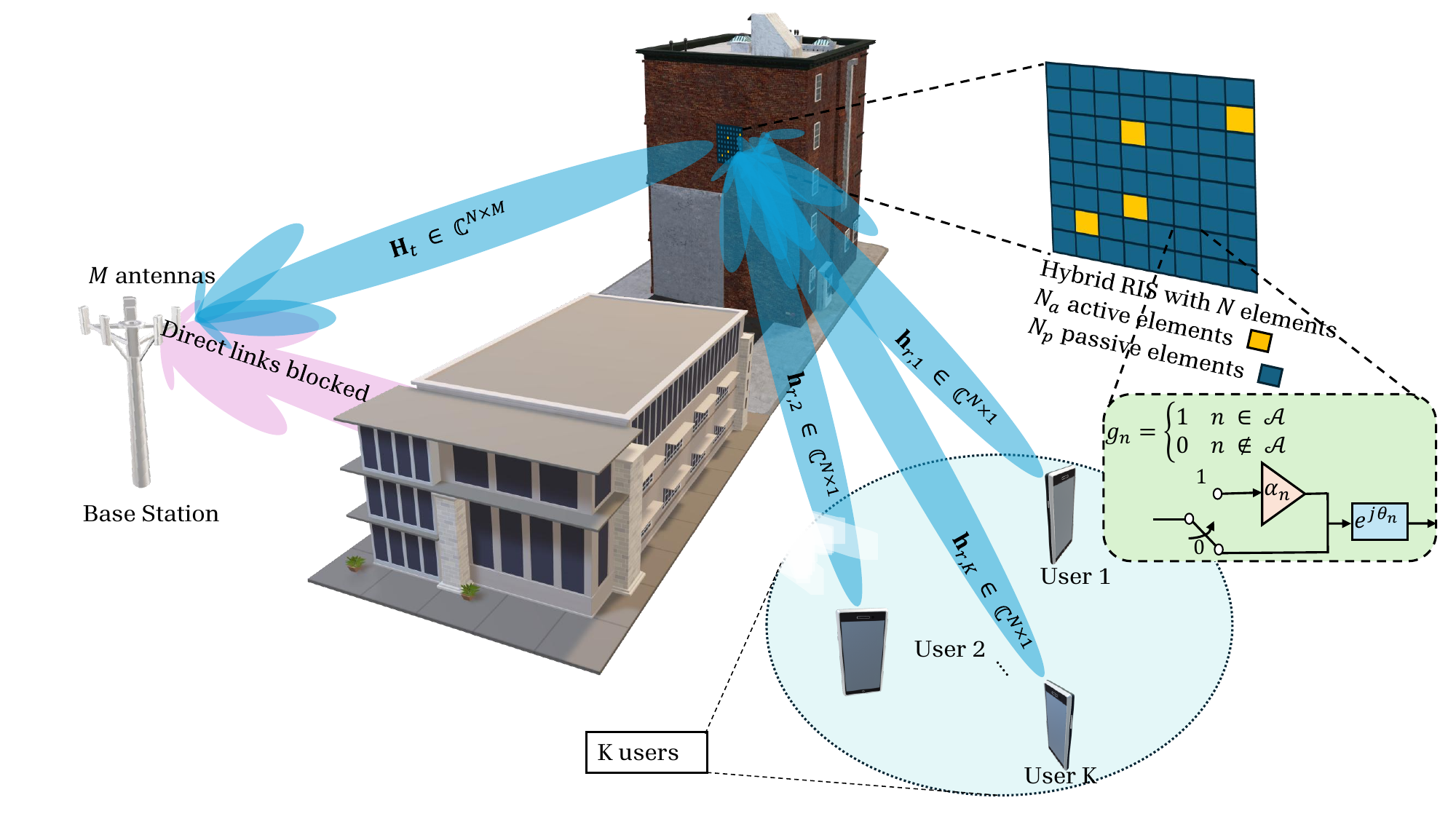}
    \caption{Considered hybrid RIS-assisted MU-MISO downlink system.}
    \label{fig:system_model}
\end{figure}

We consider a hybrid RIS-assisted MU-MISO downlink system, as illustrated in Fig.~\ref{fig:system_model}. The BS is equipped with $M$ transmit antennas and serves $K$ single-antenna users. The set of users is denoted by $\mathcal{K}=\{1,\dots,K\}$. The direct links from the BS to the users are assumed to be blocked, and communications are established through a hybrid RIS composed of $N$ elements. \blue{The hybrid RIS consists of a predefined number $N_{\mathsf{a}}$ of active elements~\cite{ju2024beamforming}, whose positions are optimized, while the remaining $N-N_{\mathsf{a}}$ elements operate in passive mode.
Let $\theta_n \in \{0,\pi\}$ denote the binary phase shift of hybrid RIS element $n$, for $n=1,\ldots,N$. Furthermore, let ${\mathcal{N} = \{1,\ldots,N\}}$ and $\mathcal{A} \subset \mathcal{N}$ denote the set of RIS elements and the subset of active elements, with $|\mathcal{N}| = N$ and $|\mathcal{A}| = N_{\mathsf{a}}$, respectively.} We denote the overall hybrid RIS coefficient matrix by the diagonal matrix $\bPhi$, 
\begin{align}
    \bPhi = \diag{\phi_1, \ldots, \phi_N},
    \label{eq:srismatrix}
\end{align}
where each diagonal element $\phi_n$ is given by
\begin{equation}
    \phi_n =
    \begin{cases}
        \alpha_n e^{j\theta_n}, & n \in \mathcal{A}~(\text{active element}), \\[4pt]
        e^{j\theta_n}, & n \notin \mathcal{A}~(\text{passive element}).
    \end{cases}
    \label{eq:hybridris}
\end{equation}
Here, $\alpha_n$ denotes the real-valued amplitude coefficient of the $n$th active element $(1\leq\alpha_n \leq \alpha^{\mathsf{max}})$, where $\alpha^{\mathsf{max}}$ represents the maximum allowable amplitude coefficient.
Based on \eqref{eq:hybridris}, we decompose $\bPhi$ into two diagonal matrices, $\bPhi_{\mathsf{a}}$ and $\bPhi_{\mathsf{p}}$, which represent the active and passive reflection coefficients, respectively, i.e.,
\begin{align}
    \bPhi 
    = \underbrace{\diag{\mathbbm{1}_N^\mathcal{A}}\bPhi}_{\triangleq~\bPhi_{\mathsf{a}}}
    + \underbrace{\big(\mI_N - \diag{\mathbbm{1}_N^\mathcal{A}}\big)\bPhi}_{\triangleq~\bPhi_{\mathsf{p}}},
\end{align}
where $\mathbbm{1}_N^\mathcal{A}$ is an indicator vector of dimension $N$ with ones at the positions corresponding to the active element set $\mathcal{A}$. This decomposition is introduced to explicitly separate the contributions of the active and passive elements in the subsequent~analysis.
Let $\vx = [x_1, \ldots, x_K]^\T \in\mathbb{C}^{K\times 1}$ be the vector of transmitted symbols at the BS, with $\mathbb{E}\{\vx\vx^\mathsf{H}\}=\mathbf{I}_K$. Furthermore, let $\mF=[\vf_1,\ldots,\vf_K]\in\mathbb{C}^{M\times K}$ be the transmit beamforming matrix. The transmitted signal from the BS is given by $\vs=\mF\vx$. The received signal at user $k$, denoted by $y_k$, can be expressed as
\begin{equation}
    y_k = \vhrkH \bPhi \mHt \mF \vx + \vhrkH \bPhi_{\mathsf{a}} \vz_{\mathsf{R}} + z_k,
    \label{eq:receive}
\end{equation}
where $\mHt \in \mathbb{C}^{N\times M}$ and $\vhrk \in \mathbb{C}^{N}$ denote the channels between the BS and the RIS and between the RIS and user $k$, respectively. We assume that perfect CSI is available at the BS.
In \eqref{eq:receive}, $z_k\sim \mathcal{CN}(0,\sigma_k^2)$ denotes the additive white Gaussian noise (AWGN) at user $k$, while $\vz_{\mathsf{R}} \sim \mathcal{CN}(0,\sigma_{\mathsf{R}}^2 \mI_N)$ is the thermal noise introduced by the active elements at the hybrid~RIS.
\\\indent 
Assume that the symbol $x_k$ is intended for user $k$. The signal-to-interference-plus-noise ratio (SINR) at this user is given by
\begin{align}
    \gamma_k = \frac{\left| \vhrkH \bPhi \mHt \vf_k \right|^2}
    {\displaystyle\sum_{i \neq k}^{K} \left| \vhrkH \bPhi \mHt \vf_i \right|^2 + \sigma_k^2 + \sigma_{\mathsf{R}}^2 \left\| \vhrkH \bPhi_{\mathsf{a}} \right\|^2 }.
    \label{eq:sinr}
\end{align}
Thus, the SE of user $k$ is given by $\mathsf{SE}_k = \log_{2} \left(1+\gamma_k\right)$, and the total SE of the system is
\begin{equation}
    \mathsf{SE} = \sum_{k\in\mathcal{K}} \mathsf{SE}_k.
    \label{eq:spectral}
\end{equation}
\subsection{Problem Formulation}
Our objective is to maximize the system EE, defined as the ratio between the SE and the total power consumption. We first present the power consumption model of the considered~system.
\subsubsection{Power Model}
The total power consumption comprises two components: 1) the power consumption at the BS and 2) the power consumption at the hybrid RIS. The overall BS power consumption is modeled~as~\cite{zhi2022active}\footnote{Notice that the users' power consumption is not included in the BS power consumption, similarly to~\cite{ntougias2025hybrid}. It can be readily added (as a constant term) in the total power consumption model.}
\begin{equation}
    P_{\mathsf{BS}} = \dfrac{1}{\xi}P_\mathsf{t} + P_{\mathsf{c,BS}},
\end{equation}
where $\xi \in (0,1]$ denotes the power amplifier efficiency, ${P_{\mathsf{t}} = \tr{\mF \mF^{\mathsf{H}}}}$ is the BS transmit power, and $P_{\mathsf{c,BS}}$ represents the static circuit power.

For the hybrid RIS with $N$ reflecting elements, each element requires a phase shifter, leading to a power consumption of $N P_{\mathsf{ps}}$, where $P_{\mathsf{ps}}$ is the power consumed by one phase shifter. Active elements further incur amplifier-related power consumption, consisting of a static bias power $P_{\mathsf{dc}}$ required to keep the amplifier active~\cite{rodrigues2023efficient}, and a dynamic power term $P_{\mathsf{r}}$, which denotes the output power of the active RIS elements and accounts for both the amplified incident signal and the amplified thermal noise 
\begin{equation}
    P_{\mathsf{r}} =
    \tr{\!
        \bPhi_{\mathsf{a}}
        \mHt \mF \mF^{\mathsf{H}} \mHt^{\mathsf{H}}
        \bPhi_{\mathsf{a}}^{\mathsf{H}}
    }
    +
    \tr{
        \sigma_{\mathsf{R}}^{2}\, \bPhi_{\mathsf{a}} \bPhi_{\mathsf{a}}^{\mathsf{H}}
    }.
\end{equation}
Combining these terms, the total hybrid RIS power consumption is given by
\begin{equation}
    P_{\mathsf{RIS}} = N P_{\mathsf{ps}} + N_{\mathsf{a}} P_{\mathsf{dc}} + \dfrac{1}{\zeta} P_{\mathsf{r}},
\end{equation}
where $\zeta \in (0,1]$ denotes the amplifier efficiency at the RIS active elements. Finally, the overall system power consumption of the considered hybrid RIS-aided system is given by
\begin{equation}
    P_{\mathsf{tot.}} = P_{\mathsf{BS}} + P_{\mathsf{RIS}} = \dfrac{1}{\xi}P_\mathsf{t} +\dfrac{1}{\zeta} P_{\mathsf{r}} + P_{\mathsf{c}} ,
    \label{eq:powertot}
\end{equation}
where $P_{\mathsf{c}}  \triangleq P_{\mathsf{c,BS}} + N P_{\mathsf{ps}} + N_{\mathsf{a}} P_{\mathsf{dc}}$ denotes the total static power consumption of the BS and RIS.
\subsubsection{EE Design Problem}
We aim to jointly design the BS beamforming matrix $\mF$, the hybrid RIS reflection matrix $\bPhi$, and the active-element set $\mathcal{A}$ to maximize the system EE under the BS and RIS power budgets, the QoS, and hardware limitations, including the amplitude-coefficient limits of the active elements, binary phase-shift control, and the fixed active-element cardinality. Mathematically, this problem is formulated as
\begin{subequations}\label{op_main_1}
    \begin{align} 
       \quad \underset{\substack{ \mF, \bPhi, \mathcal{A} }}{\textrm{maximize}} \quad &  \mathsf{EE} \triangleq \frac{ \mathsf{SE} }{ P_{\mathsf{tot.}} }\label{eq:maxeea} \\
        \textrm{subject to} \quad
        &\tr{\mF \mFh} \leq P_{\mathsf{t}}^{\mathsf{max}},
\label{eq:maxeeb} \\
        &P_{\mathsf{r}} \leq P_{\mathsf{r}}^{\mathsf{max}},
\label{eq:maxeec} \\
        &\mathsf{SE}_k \geq \mathsf{SE}^{\mathsf{min}}_k, \ \forall\, k \in \mathcal{K},
 \label{eq:maxeed} \\
       &  1\leq\alpha_n \leq \alpha^{\mathsf{max}} , \ \forall \, n \in \mathcal{A},
       \label{eq:maxeee} \\
       &\theta_n \in \{0, \pi\}, \ \forall n \in \mathcal{N},
\label{eq:maxeef}\\
        &|\mathcal{A}| = N_{\mathsf{a}},
\label{eq:maxeeg}
    \end{align}
\end{subequations}
where $P_{\mathsf{t}}^{\mathsf{max}}$ and $P_{\mathsf{r}}^{\mathsf{max}}$ are the transmit power budgets at the BS and the hybrid RIS, respectively, and $\mathsf{SE}_k^{\mathsf{min}}$ denotes the minimum SE requirement of user $k$. Constraints \eqref{eq:maxeeb} and \eqref{eq:maxeec} impose the transmit power limitations at the BS and the hybrid RIS, respectively. Constraint \eqref{eq:maxeed} ensures the QoS requirement of each user by enforcing a minimum SE level. Furthermore, constraint \eqref{eq:maxeee} restricts the amplitude coefficients of the active RIS elements to their feasible range. Constraint \eqref{eq:maxeef} imposes binary phase-shift control for each RIS element, and \eqref{eq:maxeeg} fixes the number of active elements~to~$N_{\mathsf{a}}$.

The EE maximization problem in \eqref{op_main_1} is a non-convex mixed-integer program. The main challenges arise from the fractional and non-convex objective function in \eqref{eq:maxeea}, the coupling among the beamforming and RIS design variables in both the SE and power-consumption expressions, and the discrete nature of the RIS design variables. In particular, the selection of the active-element set $\mathcal{A}$ is combinatorial, while constraints \eqref{eq:maxeec}--\eqref{eq:maxeeg} are difficult to handle due to the RIS power constraint, QoS requirements, binary phase-shift control, and active-element selection. Therefore, solving \eqref{op_main_1} optimally is highly challenging. In the next section, we present an efficient solution based on AO and deep unfolding.
\section{Proposed Deep Unfolded Alternating Optimization Method} \label{sec:deepAO}
In this section, we present our proposed solution to the EE maximization problem in \eqref{op_main_1}. We adopt an AO framework and decouple the original problem into two subproblems. The BS beamforming and power allocation subproblem, with fixed RIS parameters, is solved in closed form and detailed in Section~\ref{subsec:ao}. The hybrid RIS optimization subproblem, with fixed BS variables, is then handled using a model-driven deep unfolding method in Section~\ref{subsec:deepunfold}.

\blue{Before applying AO, we employ ZF beamforming to cancel the multiuser interference and simplify the SINR expression in \eqref{eq:sinr}. The ZF requires $M\geq K$ and the effective cascaded channel $\mH$ to have full row rank, such that $\mH^{\mathsf H}\mH$ is invertible. Under these conditions, the beamforming matrix is given by}
\begin{equation}
    \mF = \mH(\mH^{\mathsf{H}}\mH)^{-1}\mP^{\frac{1}{2}} \triangleq \mH^{\dagger} \mP^{\frac{1}{2}},
\end{equation}
where $\mH^{\mathsf{H}} \triangleq \mHrh \bPhi \mHt$ denotes the cascaded channel from the BS to the users, $\mHr = [\vh_{\mathsf{r},1}, \vh_{\mathsf{r},2}, \dots, \vh_{\mathsf{r}, K}] \in \mathbb{C}^{N \times K}$ collects the channels between the hybrid RIS and the users, and $\mH^{\dagger} \triangleq \mH(\mH^{\mathsf{H}}\mH)^{-1}$. Matrix $\mP$ is given by $\mP = \diag{p_{1},\ldots,p_{K}}$, where $p_k$ denotes the power allocation factor associated with user $k$.
In the following, we reformulate problem \eqref{op_main_1} under the ZF beamforming structure.

\subsection{Reformulation of \texorpdfstring{\eqref{op_main_1}}{(1)} with ZF Beamforming}
With ZF, the SINR of user $k$ in \eqref{eq:sinr} can be rewritten as
\begin{align}
    \gamma_k = \frac{p_k}{\sigma_k^2 + \sigma_{\mathsf{R}}^2 \norm{\vhrkH \bPhi_{\mathsf{a}}}^2} .
\end{align}
Furthermore, the BS transmit power $P_{\mathsf{t}}$ and the hybrid RIS dynamic amplifier power $P_{\mathsf{r}}$, respectively, become
\begin{equation}
    P_{\mathsf{t}} = \tr{\mP^{\frac{1}{2}}(\mH^{\mathsf{H}}\mH)^{-1}\mP^{\frac{1}{2}}}
    = \tr{\mP(\mH^{\mathsf{H}}\mH)^{-1}},
    \label{eq:zf_powerbs}
\end{equation}

\begin{align}
    P_{\mathsf{r}}(\bPhi_{\mathsf{a}}, \mP)\! &= \!
    \tr{\!
        \bPhi_{\mathsf{a}}
        \mHt \mH^{\dagger}\mP (\mH^{\dagger})^{\mathsf{H}} \mHt^{\mathsf{H}}
        \bPhi_{\mathsf{a}}^{\mathsf{H}}
    \!}\! \!+\! \tr{\!
        \sigma_{\mathsf{R}}^{2}\, \bPhi_{\mathsf{a}} \bPhi_{\mathsf{a}}^{\mathsf{H}}\!}\!.
    \label{eq:zf_powerris}
\end{align}
As a result, the system EE is now given by
\begin{align}
    \mathsf{EE} = \dfrac{ \sum_{k=1}^{K} \log_2\!\left(1 + \dfrac{p_k}{\sigma_k^2 + \sigma_{\mathsf{R}}^2 \norm{\vhrkH \bPhi_{\mathsf{a}}}^2}\right)}{\dfrac{1}{\xi}P_{\mathsf{t}} + P_{\mathsf{c}} + \dfrac{1}{\zeta}P_{\mathsf{r}}(\bPhi_{\mathsf{a}}, \mP)}. \label{eq_new_EE}
\end{align}
\\\indent
The objective in \eqref{eq_new_EE} remains challenging to optimize because $\bPhi_{\mathsf{a}}$ and $\mP$ appear in the denominators of both the SINR and EE expressions. This renders the problem fractional and non-convex, making direct optimization challenging. In particular, the dynamic amplifier power depends on the RIS coefficients in a highly nonlinear manner, while the noise introduced by the active RIS elements appears as an additional term in the SINR denominator. These coupled nonlinearities make direct optimization challenging.
\\\indent
\blue{To overcome the above challenge, we make the following observations. First, the noise power at user $k$ introduced by the active RIS elements, i.e., $\sigma_{\mathsf{R}}^2 \big\| \vhrkH \bPhi_{\mathsf{a}} \big\|^2$, is relatively small compared with $\sigma_k^2$ under the considered channel conditions, since $\sigma_{\mathsf{R}}^2$ and $\sigma_k^2$ are of comparable magnitude~\cite{zhi2022active,11176921}, while the RIS--user path loss limits $\big\| \vhrkH \bPhi_{\mathsf{a}} \big\|^2$. Thus, we neglect this term in the SINR expression to make the objective function more tractable. Second, the amplifier power at the active RIS elements in the denominator of the EE, i.e., $P_{\mathsf r}(\bPhi_{\mathsf a},\mP)$, satisfies $P_{\mathsf r}\leq P_{\mathsf r}^{\mathsf{max}}$. Replacing $P_{\mathsf r}$ with $P_{\mathsf r}^{\mathsf{max}}$ increases the EE denominator and thus provides a lower bound on the exact EE when the SE remains unchanged. The resulting approximation gap is small in the low RIS-power regime~\cite{zhi2022active}.}
\\\indent 
Based on the above observations,  we consider the following approximation of the EE, denoted by $\widetilde{\mathsf{EE}}$
\begin{align}
    \widetilde{\mathsf{EE}}
    = 
    \frac{ \sum_{k=1}^{K} \log_{2}\!\left(1 + \dfrac{p_k}{\sigma_k^2} \right) }
    { \frac{1}{\xi} P_{\mathsf{t}} + P_{\mathsf{c}} + \frac{1}{\zeta}P_{\mathsf{r}}^{\mathsf{max}} },
    \label{eq_new_EE_approx}
\end{align}
where the term 
$\sigma_{\mathsf{R}}^2 \big\| \vhrkH \bPhi_{\mathsf{a}} \big\|^2$ 
is removed from the SINR denominator, and  $P_{\mathsf{r}}(\bPhi_{\mathsf{a}}, \mP)$  is replaced by its maximum allowable value $P_{\mathsf{r}}^{\mathsf{max}}$ based on constraint~\eqref{eq:maxeec}. 
The approximation becomes tight when the RIS uses most of its available power budget for signal amplification and when the RIS power budget is relatively small compared with the BS-side power consumption. This aligns with practical systems, where RISs operate under low-power conditions. In the simulation results, we also quantify the approximation accuracy by comparing $\widetilde{\mathsf{EE}}$ with the original EE in~\eqref{eq_new_EE}.

As a result, problem \eqref{op_main_1} can be reformulated as 
\begin{subequations}
    \label{opt_ao}
    \begin{align} 
         \quad\underset{\substack{\mP, \bPhi, \mathcal{A} } }{\textrm{maximize}} \quad &  \widetilde{\mathsf{EE}}, \label{eq:maaxeea} \\
        \textrm{subject to} \quad
        &\tr{\mP(\mH^\mathsf{H}\mH)^{-1}} \leq P_{\mathsf{t}}^{\mathsf{max}},
\label{eq:maaxeeb} \\
        &P_{\mathsf{r}} \leq P_{\mathsf{r}}^{\mathsf{max}}, 
\label{eq:maaxeec}\\
        &p_k \geq \sigma_k^2 (2^{\mathsf{SE}^{\mathsf{min}}_k}-1), \ \forall k \in \mathcal{K}, %
 \label{eq:maaxeed}  \\
        & \eqref{eq:maxeee}-\eqref{eq:maxeeg}.
    \end{align}
\end{subequations}
The resulting optimization problem remains non-convex. We next present the proposed solution method.

\subsection{AO Method}
In the AO framework, the power allocation matrix $\mP$ and the hybrid RIS coefficients $(\bPhi,\mathcal{A})$ are optimized in an alternating manner by solving two corresponding subproblems while keeping the other variables fixed, as detailed below.

\subsubsection{Power Allocation Subproblem}\label{subsec:ao}
For a given hybrid RIS coefficient matrix $\bPhi$ and active set $\mathcal{A}$, problem \eqref{opt_ao} can be rewritten as
\begin{subequations}
    \label{opt_beampattern}
    \begin{align} 
         \quad \underset{\substack{ \mP }}{\textrm{maximize}} \quad &  \dfrac{ \sum_{k=1}^{K} \log_2\left(1 + \dfrac{p_k}{\sigma_k^2}\right)}{\bar{P} + \dfrac{1}{\xi} \sum_{k=1}^{K} p_k t_k }, \label{eq:pmaxeea} \\
        \textrm{subject to} \quad
        &\sum_{k=1}^{K} p_k t_k \leq P_{\mathsf{t}}^{\mathsf{max}},
\label{eq:pmaxeeb} \\
    &\sum_{k=1}^{K} p_k d_k
    +
    \tr{
        \sigma_{\mathsf{R}}^{2}\, \bPhi_{\mathsf{a}} \bPhi_{\mathsf{a}}^{\mathsf{H}}} \leq P_{\mathsf{r}}^{\mathsf{max}}, 
\label{eq:pmaxeec}\\
        &p_k \geq p_k^{\mathsf{min}}, \ \forall k \in \mathcal{K},
\label{eq:pmaxeed} 
    \end{align}
\end{subequations}
where $\bar{P} \triangleq P_{\mathsf{c}} + \frac{1}{\zeta}P_{\mathsf{r}}^{\mathsf{max}}$ and $p_k^{\mathsf{min}} \triangleq \sigma_k^2 (2^{\mathsf{SE}^{\mathsf{min}}_k}-1)$ are constants. Furthermore, $t_k$ and $d_k$ are the $k$th diagonal elements of $(\mH^{\mathsf{H}} \mH)^{-1}$ and $(\mH^{\dagger})^{\mathsf{H}} \mHt^{\mathsf{H}} \bPhi_{\mathsf{a}}^{\mathsf{H}} \bPhi_{\mathsf{a}} \mHt \mH^{\dagger}$, respectively.

Although all constraints \eqref{eq:pmaxeeb}--\eqref{eq:pmaxeed} are convex, the objective in \eqref{eq:pmaxeea} is fractional and hence non-concave. To solve problem \eqref{opt_beampattern}, we employ the Dinkelbach method~\cite{dinkelbach1967nonlinear}, which iteratively solves the following problem in iteration $(i+1)$    
\begin{align}   \label{power_alo}
          \underset{\substack{\mP}}{\textrm{maximize}}~  f(\mP),\
        \textrm{subject to}~ \eqref{eq:pmaxeeb}- \eqref{eq:pmaxeed},
    \end{align}
where $f(\mP)\!\triangleq\!\!\sum\limits_{k=1}^{K} \log_2\left(1 + \dfrac{p_k}{\sigma_k^2}\right)\!-\! \lambda^{(i)}\!\left(\!\bar{P} \!+ \!\dfrac{1}{\xi}\sum\limits_{k=1}^{K} p_k t_k \!\right)$ and
\begin{align}
\lambda^{(i)} \triangleq \dfrac{\sum_{k=1}^{K} \log_2\left(1 + \dfrac{p_k^{(i)}}{\sigma_k^2}\right)}{\bar{P} + \dfrac{1}{\xi}\sum_{k=1}^{K} p_k^{(i)} t_k} \label{eq_update_lambda}
\end{align}
denotes the Dinkelbach parameter, which is updated from the solution of the problem in the previous iteration $i$, i.e., ${ {\mathbf{P}^{(i)}}^* =[p_k^{(i)}] }$. Problem \eqref{power_alo} is convex, and its closed-form solution is derived as follows. 

The Lagrangian of \eqref{power_alo} can be written as 
\begin{align} \label{eq:lagrangian}
\mathcal{L}(\mP,\bmu) &=
\mu_0 \!\left(\sum_{k=1}^{K} p_k t_k - P_{\mathsf{t}}^{\mathsf{max}}\right)- \sum_{k=1}^{K} \mu_k \,(p_k - p_k^{\mathsf{min}}) 
\nonumber\\
&\quad+ \mu_{K+1} \!\left(\sum_{k=1}^{K} p_k d_k - \hat{P}_{\mathsf{r}}^{\mathsf{max}}\right) - f(\mP) .
\end{align}
where $\hat{P}_{\mathsf{r}}^{\mathsf{max}} = P_{\mathsf{r}}^{\mathsf{max}} - 
\mathrm{tr}\!\left(\sigma_{\mathsf{R}}^{2}\, 
\bPhi_{\mathsf{a}} \bPhi_{\mathsf{a}}^{\mathsf{H}}\right)$ 
is introduced for notational simplicity and $\{\mu_k\}_{k=0}^{K+1}$ are the dual variables, $\mu_k \ge 0, \forall k$. Then, the KKT conditions of \eqref{eq:lagrangian} include
\begin{subequations}
    \begin{align}
    &g(\mu_0, \mu_{K+1}) - \mu_k = \frac{1}{(p_k+\sigma_k^{2})\ln 2},\ \forall k\in\mathcal{K},
    \label{eq:kkta}
    \\
    &\mu_0 \!\left(\sum_{k=1}^{K} p_k t_k - P_{\mathsf{t}}^{\mathsf{max}}\right) = 0,
    \label{eq:kktb}
    \\
    &\mu_{K+1} \!\left(\sum_{k=1}^{K} p_k d_k - \hat{P}_{\mathsf{r}}^{\mathsf{max}}\right) = 0,
    \label{eq:kktc}
    \\
    &\mu_k \,(p_k - p_k^{\mathsf{min}}) = 0,\quad \forall k\in\mathcal{K},
    \label{eq:kktd}
    \end{align}
\end{subequations}
where $g(\mu_0, \mu_{K+1}) \triangleq \left(\dfrac{\lambda^{(i)}}{\xi}+\mu_0\right)t_k + \mu_{K+1} d_k$. By substituting \eqref{eq:kktd} into \eqref{eq:kkta}, we obtain
\begin{equation}
    p_k \;=\; \max\!\left\{ \frac{1}{g(\mu_0, \mu_{K+1}) \ln 2}\;-\;\sigma_k^{2},\;\; p_k^{\mathsf{min}}\right\}.
    \label{eq:pallkkt}
\end{equation}
It is observed from \eqref{eq:pallkkt} that $p_k$ is a function of the dual variables
$\mu_0$ and $\mu_{K+1}$, together with the complementary slackness conditions
in \eqref{eq:kktb}--\eqref{eq:kktc}, the optimal dual variables $\mu_0$ and
$\mu_{K+1}$ are determined such that the following conditions are satisfied
\begin{subequations}\label{61}
\begin{align}
&\sum_{k=1}^K t_k\, p_k(\mu_0,\mu_{K+1})
\le P_t^{\mathsf{max}}, \label{61a}\\&
\sum_{k=1}^K d_k\, p_k(\mu_0,\mu_{K+1})
\le \hat P_r^{\mathsf{max}}, \label{61b}\\&
\mu_0\!\left(\sum_{k=1}^K t_k\, p_k(\mu_0,\mu_{K+1})-P_t^{\mathsf{max}}\right)=0, \label{61c}\\&
\mu_{K+1}\!\left(\sum_{k=1}^K d_k\, p_k(\mu_0,\mu_{K+1})-\hat P_r^{\mathsf{max}}\right)=0, \label{61d}
\end{align}
\end{subequations}
Since $p_k(\mu_0,\mu_{K+1})$ is monotonically non-increasing with respect to
both $\mu_0$ and $\mu_{K+1}$, the constraints in \eqref{61a} and \eqref{61b}
define monotone functions of the dual variables. Therefore, the optimal
$(\mu_0^\star,\mu_{K+1}^\star)$ can be efficiently obtained via a
bisection search, where for a fixed $\mu_{K+1}$, the value of $\mu_0$ is
uniquely determined such that \eqref{61a} holds with equality if active,
and $\mu_{K+1}$ is subsequently updated to satisfy \eqref{61b}. This procedure is repeated iteratively until convergence,
as summarized in~Algorithm~\ref{alg:power_allo}. 

\begin{algorithm}[t]\small
\caption{\small Dinkelbach Algorithm for Solving Power Allocation Subproblem~\eqref{power_alo}}
\label{alg:power_allo}
\begin{algorithmic}[1]
    \Require $\mH$ and $\bPhi_{\mathsf{a}}$
    \State Obtain $t_k$ and $d_k$ as the $k$-th diagonal elements of $(\mH^{\mathsf{H}} \mH)^{-1}$ and $(\mH^{\dagger})^{\mathsf{H}} \mHt^{\mathsf{H}} \bPhi_{\mathsf{a}}^{\mathsf{H}} \bPhi_{\mathsf{a}} \mHt \mH^{\dagger}$, respectively. 
    \State Initialize $\lambda^{(0)} > 0$.
    \For{$i = 1, \ldots, I$}
        \State Find $\mu_0^\star, \mu_{K+1}^\star$ satisfying \eqref{61a}, \eqref{61b} via bisection search. 

        \State Update power allocation $p_k^{(i)} \leftarrow p_k(\mu_0^\star, \mu_{K+1}^\star), \forall k$

        \State Update Dinkelbach factor based on \eqref{eq_update_lambda}.

    \EndFor
    \State \Return $\{p_k^\star\} \leftarrow \{p_k^{(I)}\}$
\end{algorithmic}
\end{algorithm}

\subsubsection{RIS Optimization Subproblem} \label{subsec:deepunfold}
With $\mP$ fixed in problem \eqref{opt_ao}, we aim to solve for $\bPhi$ and $\mathcal{A}$. Note that the numerator of the objective function is independent of $\bPhi$ and $\mathcal{A}$, and only the BS transmit power in \eqref{eq:zf_powerbs} depends on $\bPhi$ and $\mathcal{A}$, while other power consumption components remain constant. Based on these observations, the hybrid RIS optimization in \eqref{opt_ao} can be formulated as
\begin{subequations}
    \label{ris_opt}
    \begin{align}
        \quad \underset{\substack{\bPhi, \mathcal{A}}}{\textrm{minimize}} \quad &  \rho(\bPhi) \triangleq \tr{\mP(\mH^{\mathsf{H}}\mH)^{-1}},  \label{eq:beama} \\
        \textrm{subject to} \quad
        &\tr{\mP(\mH^{\mathsf{H}}\mH)^{-1}} \leq P_{\mathsf{t}}^{\mathsf{max}},
\label{eq:beamb}  \\
        &P_{\mathsf{r}} \leq P_{\mathsf{r}}^{\mathsf{max}}, 
\label{eq:beamc} \\
        &\eqref{eq:maxeee}-\eqref{eq:maxeeg}.
\label{eq:beamd}
    \end{align}
\end{subequations}

The resulting problem involves combinatorial active-element selection and nonconvex constraints, making exact optimization computationally challenging, particularly for large RIS. We propose an efficient solution to this problem by leveraging PGD and deep unfolding.
\textcolor{blue}{
\paragraph{PGD Framework}
To enable gradient-based optimization, we first relax the discrete RIS constraints by assuming that all RIS elements are active, i.e., $\mathcal{A}\equiv\mathcal{N}$, and allowing $\theta_n\in[0,2\pi)$ in~\eqref{eq:hybridris}. At iteration $j$, the RIS coefficients are updated as
\begin{equation}
    \tilde{\bPhi}^{(j)}
    =
    \bPhi^{(j)}
    -
    \eta_{\bPhi}^{(j)}
    \nabla_{\bPhi}\rho\!\left(\bPhi^{(j)}\right),
    \label{eq:pgdris}
\end{equation}
where $\eta_{\bPhi}^{(j)}$ denotes the step size. Under this relaxation, the gradient is given by
\begin{equation}
    \nabla_{\bPhi}\rho(\bPhi)
    =
    -2\left( \mHr (\mH^{\mathsf{H}}\mH)^{-1} \mP
    (\mH^{\dagger})^{\mathsf{H}}
    \mHt^{\mathsf{H}} \right) \circ \mI_N,
    \label{eq:gradientris}
\end{equation}
where the Hadamard product $(\circ)$ with $\mI_N$ extracts the diagonal entries associated with the RIS coefficients \cite{petersen2008matrix}.
\\
The update in \eqref{eq:pgdris} does not necessarily satisfy the constraints in~\eqref{ris_opt}. We therefore define the feasible set of this problem~as
\begin{equation}
    \begin{aligned}
    \mathcal{F}
    \triangleq
    \big\{(\bPhi,\mathcal{A}) :\;&
    \rho(\bPhi)
    \leq
    P_{\mathsf{t}}^{\mathsf{max}},
    \;
    P_{\mathsf{r}}(\bPhi)
    \leq
    P_{\mathsf{r}}^{\mathsf{max}},
    \\[-1mm]
    &
    \eqref{eq:maxeee}-\eqref{eq:maxeeg}
    \big\}.
    \end{aligned}
    \label{eq:feasible_set}
\end{equation}
Since the feasible set $\mathcal{F}$ involves discrete RIS constraints and coupled power constraints, we adopt a low-complexity projection-based procedure. \footnote{The exact Euclidean projection problem over $\mathcal{F}$ is non-convex and computationally challenging due to the discrete RIS constraints and coupled power constraints.}
\\
We, therefore, use a feasibility-preserving projected-gradient procedure. Since binary phase shifts restrict feasible RIS coefficients to the real axis, the Euclidean projection depends only on the real part, while the imaginary part contributes a constant to the squared distance. First, the phase of each RIS coefficient is quantized to the binary set $\{0,\pi\}$ to satisfy the discrete phase-shift constraint in \eqref{eq:maxeef}. Next, the active set $\mathcal{A}$ is determined by selecting the $N_{\mathsf{a}}$ elements with the largest absolute real parts in accordance with \eqref{eq:maxeeg}, while the remaining elements are assigned as passive. Subsequently, the amplitudes of the selected active elements are clipped to the feasible interval defined by \eqref{eq:maxeee}. The cascaded channel is then recomputed, and the RIS and BS power constraints in \eqref{eq:beamc} and \eqref{eq:beamb}, respectively, are explicitly checked. If either power constraint is violated, the candidate is rejected, and the previous feasible iterate is retained. The detailed implementation is presented~below.
\paragraph{Deep Unfolded PGD}
In the PGD method, the step-size sequence $\{\eta_{\bPhi}^{(j)}\}_{j=0}^{J-1}$ critically affects the convergence behavior and the quality of the obtained solution. Although adaptive schemes such as line search or backtracking~\cite{boyd2004convex} can select suitable step sizes, they introduce additional computations at each iteration. To avoid this overhead, we employ a model-driven deep unfolding approach that learns the step sizes from data.
\\
In deep unfolding, the $J$ PGD iterations are mapped onto $J$ network layers, each following the gradient update in \eqref{eq:pgdris}~\cite{hershey2014deep,deka2025comprehensive}. This transformation converts a conventional fixed-step iterative solver into a trainable architecture that can be interpreted as a specialized DNN. The resulting network preserves the structure and interpretability of the original algorithm while introducing learnable step sizes to improve optimization~performance.
\\
However, the discrete RIS constraints, including binary phase shifts, amplitude coefficient limits, and active/passive element selection~\eqref{eq:beamd}, involve non-differentiable operations that prevent direct gradient backpropagation during training. Therefore, we distinguish between training and inference. During training, these discrete constraints are replaced by smooth relaxations to enable gradient-based learning of the step sizes, while the RIS power constraint is incorporated into the training loss through a differentiable penalty. During inference, the learned step sizes are fixed, and the smooth relaxations are replaced by hard projection and explicit feasibility checks to enforce the original constraints of problem~\eqref{ris_opt}.
\\
\emph{1) Training Phase:}
Let
\begin{equation}
    \tilde{\phi}_n^{(j)}
    =
    \left[\tilde{\bPhi}^{(j)}\right]_{n,n}
    =
    \tilde{\alpha}_n^{(j)}
    e^{\mathsf{j}\tilde{\theta}_n^{(j)}},
    \qquad n\in\mathcal{N},
    \label{eq:intermediate_phi}
\end{equation}
denote the RIS coefficient after the gradient update~\eqref{eq:pgdris}. The binary phase-shift constraint is approximated as
\begin{equation}
    \bar{\theta}_n^{(j)}
    =
    -\frac{\pi}{2}
    \tanh\!\left(
    \beta_{\theta}
    \cos\!\left(
    \tilde{\theta}_n^{(j)}
    \right)
    \right)
    +
    \frac{\pi}{2},
    \qquad n\in\mathcal{N},
    \label{eq:activation}
\end{equation}
where $\beta_{\theta}>0$ controls the sharpness of the approximation.
\\
To select a predefined number $N_{\mathsf a}$ of active RIS elements under constraint~\eqref{eq:maxeeg}, we define the activity score of element $n$ as
\begin{equation}
    u_n^{(j)}
    =
    \left|
    \Re\left\{
    \tilde{\phi}_n^{(j)}
    \right\}
    \right|,
    \qquad n\in\mathcal{N}.
    \label{eq:active_score}
\end{equation}
Let $u_{(1)}^{(j)}\geq\cdots\geq u_{(N)}^{(j)}$ denote the activity scores in descending order. The selection threshold is
\begin{equation}
    \tau^{(j)}
    =
    \frac{
    u_{(N_{\mathsf{a}})}^{(j)}
    +
    u_{(N_{\mathsf{a}}+1)}^{(j)}
    }{2},
    \label{eq:selection_threshold}
\end{equation}
and the corresponding soft activity variable is
\begin{equation}
    g_n^{(j)}
    =
    \sigma\!\left(
    \beta_{\mathsf{a}}
    \left(
    u_n^{(j)}-\tau^{(j)}
    \right)
    \right),
    \qquad
    0<g_n^{(j)}<1,
    \label{eq:soft_selection}
\end{equation}
where $\sigma(x)=1/(1+e^{-x})$ is the sigmoid function and ${\beta_{\mathsf a}>0}$ controls the sharpness of the active/passive transition. As $\beta_{\mathsf a}$ increases, $g_n^{(j)}\simeq 1$ for the $N_{\mathsf a}$ elements with the largest activity scores, indicating active mode, while $g_n^{(j)}\simeq 0$ for the remaining elements, indicating passive mode.
\\
The amplitude coefficient is smoothly bounded as
\begin{equation}
    \bar{\alpha}_n^{(j)}
    =
    1 + 
    \operatorname{softplus}_{\beta_{\mathsf b}}
    \left(
    u_n^{(j)}-1
    \right)
    -
    \operatorname{softplus}_{\beta_{\mathsf b}}
    \left(
    u_n^{(j)}-\alpha^{\mathsf{max}}
    \right),
    \label{eq:soft_amplitude}
\end{equation}
where $\operatorname{softplus}_{\beta_{\mathsf b}}(x)=\beta_{\mathsf b}^{-1}\ln\!\left(1+e^{\beta_{\mathsf b}x}\right)$. As $\beta_{\mathsf b}$ increases, this smooth approximation approaches $\max\{1,\min\{u_n^{(j)},\alpha^{\mathsf{max}}\}\}$.
The relaxed amplitude and RIS coefficient are then given by
\begin{align}
    \alpha_n^{(j+1)}
    &=
    1+
    g_n^{(j)}
    \left(
    \bar{\alpha}_n^{(j)}-1
    \right),
    \label{eq:output_y}\\
    \phi_n^{(j+1)}
    &=
    \alpha_n^{(j+1)}
    e^{\mathsf{j}\bar{\theta}_n^{(j)}}.
    \label{eq:training_phi}
\end{align}
\\
The desired active-element cardinality is encouraged through
\begin{equation}
    \mathcal{L}_{\mathsf A}
    =
    \left(
    \frac{
    \sum_{n=1}^{N}g_n^{(J-1)}-N_{\mathsf a}
    }{
    N_{\mathsf a}
    }
    \right)^2,
    \label{eq:cardinality_loss}
\end{equation}
while the RIS power constraint is incorporated as
\begin{equation}
    \mathcal{L}_{\mathsf r}
    =
    \left[
    \operatorname{ReLU}
    \left(
    \frac{
    P_{\mathsf r}\!\left(\bPhi^{(J)}\right)
    }{
    P_{\mathsf r}^{\mathsf{max}}
    }
    -1
    \right)
    \right]^2.
    \label{eq:ris_constraint_loss}
\end{equation}
Accordingly, the training loss is
\begin{equation}
    \mathcal{L}
    \left(
    \boldsymbol{\eta}_{\bPhi}
    \right)
    =
    \frac{
    \rho\!\left(\bPhi^{(J)}\right)
    }{
    P_{\mathsf t}^{\mathsf{max}}
    }
    +
    \lambda_{\mathsf r}\mathcal{L}_{\mathsf r}
    +
    \lambda_{\mathsf A}\mathcal{L}_{\mathsf A},
    \label{eq:training_loss}
\end{equation}
where $\lambda_{\mathsf r}>0$ and $\lambda_{\mathsf A}>0$ are penalty weights. The terms in \eqref{eq:cardinality_loss}--\eqref{eq:training_loss} are normalized by their corresponding reference values, i.e., $N_{\mathsf a}$, $P_{\mathsf r}^{\mathsf{max}}$, and $P_{\mathsf t}^{\mathsf{max}}$, to obtain dimensionless quantities with comparable scales and improve the numerical stability of training. Since $\rho(\bPhi)$ represents the BS transmit-power term in problem~\eqref{ris_opt}, no separate penalty is required for the BS power constraint. The loss is backpropagated through all $J$ layers to learn $\boldsymbol{\eta}_{\bPhi}$.
\\
\emph{2) Inference Phase:}
Once offline training is complete, the learned step sizes are fixed. At each layer $j$, the RIS coefficients are updated according to \eqref{eq:pgdris}. The resulting coefficients are then subjected to hard projection and explicit feasibility checks to ensure the constraints of problem~\eqref{ris_opt} are satisfied.
\\
Let $z_n=[\tilde{\bPhi}^{(j)}]_{n,n}$. Since the feasible phase shifts are restricted to $\{0,\pi\}$, the corresponding RIS coefficients lie on the real axis. Thus, the projected value is $\Re\{z_n\}$; its sign determines the binary phase shift and its absolute value provides the activity score. Accordingly, the binary phase shift~is
\begin{equation}
    \hat{\theta}_n
    =
    \begin{cases}
    0, & \Re\{z_n\}\geq0,\\
    \pi, & \Re\{z_n\}<0,
    \end{cases}
    \qquad n\in\mathcal{N},
    \label{eq:hard_phase}
\end{equation}
and the activity score is
\begin{equation}
    u_n
    =
    |\Re\{z_n\}|.
    \label{eq:test_activity_score}
\end{equation}
The active set is selected as
\begin{equation}
    \hat{\mathcal{A}}
    =
    \operatorname{Top}_{N_{\mathsf{a}}}
    \left(
    \{u_n\}_{n=1}^{N}
    \right).
    \label{eq:hard_active_selection}
\end{equation}
For the selected elements,
\begin{equation}
    \hat{\alpha}_n
    =
    \max\{1, \min\{u_n,\alpha^{\mathsf{max}}\}\},
    \quad
    n\in\hat{\mathcal{A}}.
    \label{eq:hard_amplitude}
\end{equation}
The candidate RIS coefficients are then given by
\begin{equation}
    \hat{\phi}_n
    =
    \begin{cases}
    \hat{\alpha}_n e^{j\hat{\theta}_n}, & n\in\hat{\mathcal{A}},\\
    e^{j\hat{\theta}_n}, & n\notin\hat{\mathcal{A}},
    \end{cases}
    \label{eq:hard_phi}
\end{equation}
with
\begin{equation}
    \hat{\bPhi}
    =
    \diag{
    \hat{\phi}_1,\ldots,
    \hat{\phi}_N
    }.
    \label{eq:hard_Phi}
\end{equation}
Thus, the candidate RIS configuration satisfies the binary phase-shift, amplitude-coefficient, and active-element cardinality constraints.
\\
The cascaded channel is then recomputed, and the resulting BS and RIS powers are evaluated. The candidate is accepted~if
\begin{subequations}
    \label{eq:inference_feasibility}
    \begin{align}
    P_{\mathsf r}\!\left(\hat{\bPhi}\right)
    &\leq P_{\mathsf r}^{\mathsf{max}},
    \label{eq:rispowerfeasible}
    \\
    \rho\!\left(\hat{\bPhi}\right)
    &\leq \rho\!\left(\bPhi^{(j)}\right).
    \label{eq:bspowerfeasible}
    \end{align}
\end{subequations}
Since $\bPhi^{(j)}$ is feasible, the second condition also preserves the BS transmit-power constraint. Otherwise, the previous feasible iterate is retained. Since the initial iterate is feasible, feasibility is maintained throughout the inference procedure. During training, the penalty terms encourage the learned updates to satisfy the RIS power and active-element requirements.}

\begin{algorithm}[t] \small
\caption{\small Proposed Deep Unfolded PGD for Hybrid RIS Optimization
Subproblem~\eqref{ris_opt}}
\label{alg:deep_unfolding_hr_ris}
\begin{algorithmic}[1]
\Require $\mHt$, $\mHr$, $\mP$, trained step sizes
$\{\eta_{\bPhi}^{(j)}\}_{j=0}^{J-1}$, and feasible
$(\bPhi^{(0)},\mathcal{A}^{(0)})$

\For{$j=0,\ldots,J-1$}

    \State
    $\tilde{\bPhi}
    \leftarrow
    \bPhi^{(j)}-\eta_{\bPhi}^{(j)}\nabla_{\bPhi}\rho(\bPhi^{(j)})$

    \State Obtain $\{\hat{\theta},u_n\}$ and
    $\hat{\mathcal{A}}$ using
    \eqref{eq:hard_phase}--\eqref{eq:hard_active_selection}

    \State Construct $\hat{\bPhi}$ using
    \eqref{eq:hard_amplitude}--\eqref{eq:hard_Phi}

    \State Recompute the cascaded channel and evaluate
    $P_{\mathsf{t}}(\hat{\bPhi})$ and
    $P_{\mathsf{r}}(\hat{\bPhi})$

    \If{\eqref{eq:inference_feasibility} holds}
        \State
        $(\bPhi^{(j+1)},\mathcal{A}^{(j+1)})
        \leftarrow
        (\hat{\bPhi}, 
        \hat{\mathcal{A}})$
    \Else
        \State
        $(\bPhi^{(j+1)},\mathcal{A}^{(j+1)})
        \leftarrow
        (\bPhi^{(j)},\mathcal{A}^{(j)})$
    \EndIf

\EndFor

\State \Return
$\bPhi^{*}=\bPhi^{(J)}$ and
$\mathcal{A}^{*}=\mathcal{A}^{(J)}$

\end{algorithmic}
\end{algorithm}

\subsection{Overall Algorithm and Complexity Analysis}
\begin{figure*}[t]
\vspace{-0.3cm}
\centering
\begingroup
\providecommand{\bPhi}{\boldsymbol{\Phi}}
\providecommand{\mP}{\mathbf{P}}
\resizebox{0.80\textwidth}{!}{%
\begin{tikzpicture}[
  >=Latex, font=\normalsize,
  flow/.style={-Latex,semithick,draw=black!85},
  zoom/.style={densely dashed,semithick,draw=black!55},
  pa/.style={draw=blue!60!black,semithick,fill=blue!5,rounded corners=2pt,
             align=center,minimum width=20mm,minimum height=12mm,inner sep=3pt},
  ris/.style={draw=teal!65!black,semithick,fill=teal!6,rounded corners=2pt,
              align=center,minimum width=22mm,minimum height=12mm,inner sep=3pt},
  layer/.style={draw=teal!65!black,semithick,fill=white,rounded corners=2pt,
                align=center,minimum width=24mm,minimum height=11mm,inner sep=3pt},
  stage/.style={draw=teal!65!black,semithick,fill=white,rounded corners=2pt,
                align=center,minimum height=12mm,inner sep=3pt},
  operation/.style={draw=black!75,semithick,circle,fill=white,inner sep=1pt,minimum size=6mm},
  groupAO/.style={draw=black!65,densely dotted,semithick,rounded corners=6pt,inner xsep=4mm,inner ysep=5mm},
  groupRIS/.style={draw=teal!65!black,densely dotted,semithick,fill=teal!2,rounded corners=6pt,inner xsep=4mm,inner ysep=7mm},
  captionnode/.style={font=\normalsize\bfseries,fill=white,inner xsep=3pt,inner ysep=1pt},
  lab/.style={font=\small,fill=white,inner sep=1pt,align=center},
]

\node[pa] (pa1) at (1.8,10.8) {ZF and\\power allocation};
\node[ris] (ris1) at (5.0,10.8) {RIS optimization};
\node[font=\Large] (dots1) at (7.5,10.8) {$\cdots$};
\node[pa] (pas) at (10.1,10.8) {ZF and\\power allocation};
\node[ris] (riss) at (13.3,10.8) {RIS optimization};
\node[font=\Large] (dots2) at (15.8,10.8) {$\cdots$};
\node[pa] (paS) at (18.4,10.8) {ZF and\\power allocation};
\node[ris] (risS) at (21.6,10.8) {RIS optimization};

\begin{scope}[on background layer]
  \node[groupAO,fit=(pa1)(ris1)] (G1) {};
  \node[groupAO,fit=(pas)(riss)] (Gs) {};
  \node[groupAO,fit=(paS)(risS)] (GS) {};
\end{scope}
\node[captionnode] at ([yshift=1mm]G1.north) {AO iteration $1$};
\node[captionnode] at ([yshift=1mm]Gs.north) {AO iteration $s$};
\node[captionnode] at ([yshift=1mm]GS.north) {AO iteration $S$};

\node[lab] (initAO) at (-0.6,10.8) {Initial\\$\big(\bPhi^{(0)},\mathcal A^{(0)}\big)$};
\draw[flow] (initAO.east) -- (pa1.west);
\draw[flow] (pa1.east) -- node[lab,above,yshift=1mm] {$\mP^{(1)}$} (ris1.west);
\draw[flow] (ris1.east) -- (dots1.west);
\draw[flow] (dots1.east) -- (pas.west);
\draw[flow] (pas.east) -- node[lab,above,yshift=1mm] {$\mP^{(s)}$} (riss.west);
\draw[flow] (riss.east) -- (dots2.west);
\draw[flow] (dots2.east) -- (paS.west);
\draw[flow] (paS.east) -- node[lab,above,yshift=1mm] {$\mP^{(S)}$} (risS.west);
\draw[flow] (risS.east) -- (23.5,10.8)
  node[lab,right] {$\mP^{(S)},\bPhi^{(S)},\mathcal A^{(S)}$};

\node[lab] (bIn) at (0.4,7.9) {$\big(\bPhi^{(0)},\mathcal A^{(0)}\big)$};
\node[layer] (L1) at (3.1,7.9) {Unfolded\\layer $1$};
\node[font=\Large] (Ldots1) at (6.3,7.9) {$\cdots$};
\node[layer] (Lj) at (10.3,7.9) {Unfolded\\layer $j+1$};
\node[font=\Large] (Ldots2) at (14.3,7.9) {$\cdots$};
\node[layer] (LJ) at (18.0,7.9) {Unfolded\\layer $J$};
\node[lab] (bOut) at (21.5,7.9) {$\big(\bPhi^{(J)},\mathcal A^{(J)}\big)$};
\begin{scope}[on background layer]
  \node[groupRIS,fit=(bIn)(L1)(Ldots1)(Lj)(Ldots2)(LJ)(bOut)] (Ginner) {};
\end{scope}
\node[captionnode] at ([yshift=1mm]Ginner.north)
 {RIS optimization within AO iteration $s$ ($\mP^{(s)}$ fixed)};
\draw[flow] (bIn) -- (L1);
\draw[flow] (L1) -- (Ldots1);
\draw[flow] (Ldots1) -- (Lj);
\draw[flow] (Lj) -- (Ldots2);
\draw[flow] (Ldots2) -- (LJ);
\draw[flow] (LJ) -- (bOut);
\draw[zoom] (riss.south west) -- (Ginner.north west);
\draw[zoom] (riss.south east) -- (Ginner.north east);

\node[lab] (cIn) at (0.45,5.0) {$\bPhi^{(j)}$};
\node[stage,minimum width=36mm] (gradient) at (4.1,5.0)
 {Compute $\nabla_{\bPhi}\rho\big(\bPhi^{(j)}\big)$};
\node[operation] (gain) at (7.1,5.0) {$\times$};
\node[operation] (sum) at (9.4,5.0) {$+$};
\node[stage,minimum width=30mm] (trial) at (12.9,5.0)
 {$\widetilde{\bPhi}^{(j)}$\\(gradient step)};
\node[stage,minimum width=45mm] (proj) at (17.4,5.0)
 {Hard projection\\and feasibility checks};
\node[lab] (cOut) at (21.9,5.0)
 {$\big(\bPhi^{(j+1)},\mathcal A^{(j+1)}\big)$};
\begin{scope}[on background layer]
  \node[groupRIS,fit=(cIn)(gradient)(gain)(sum)(trial)(proj)(cOut)] (Gdetail) {};
\end{scope}
\node[captionnode] at ([yshift=1mm]Gdetail.north)
 {One unfolded RIS layer (inference)};
\draw[flow] (cIn) -- (gradient);
\draw[flow] (gradient) -- (gain);
\draw[flow] (7.1,5.8) -- node[lab,left] {$-\eta_{\bPhi}^{(j)}$} (gain.north);
\draw[flow] (gain) -- (sum);
\draw[flow] (sum) -- (trial);
\draw[flow] (trial) -- (proj);
\draw[flow] (proj) -- (cOut);
\draw[flow] (cIn.south) |- (9.4,4.34) -- (sum.south);
\draw[zoom] (Lj.south west) -- (Gdetail.north west);
\draw[zoom] (Lj.south east) -- (Gdetail.north east);

\node[lab] (dIn) at (0.45,1.62)
 {$z_n=[\widetilde{\bPhi}^{(j)}]_{n,n}$};
\node[stage,minimum width=38mm] (phase) at (4.4,2.15)
 {Binary phase\\$\hat{\theta}_n\in\{0,\pi\}$};
\node[stage,minimum width=38mm] (score) at (4.4,0.55)
 {Activity score\\$u_n=|\operatorname{Re}\{z_n\}|$};
\node[stage,minimum width=38mm] (select) at (8.7,0.55)
 {$\hat{\mathcal A}=\mathrm{Top}_{N_{\mathsf a}}\{u_n\}$};
\node[stage,minimum width=38mm] (amp) at (13.45,0.55)
 {$\hat{\alpha}_n=\max\{1, \min\{u_n,\alpha^{\mathsf{max}}\}\}$\\for $n\in\hat{\mathcal A}$};
\node[stage,minimum width=41mm] (construct) at (13.45,2.15)
 {Build $\hat{\bPhi}$\\active: $\hat{\alpha}_n e^{j\hat{\theta}_n}$; passive: $e^{j\hat{\theta}_n}$};
\node[stage,minimum width=39mm] (check) at (18.15,2.15)
 {$P_r(\hat{\bPhi})\leq P_r^{\mathsf{max}}$\\$\rho(\hat{\bPhi})\leq\rho(\bPhi^{(j)})$};
\node[stage,minimum width=38mm] (accept) at (22.65,2.15)
 {If both hold: $\begin{cases}
     \bPhi^{(j+1)} = \hat{\bPhi}\\
     \mathcal A^{(j+1)}=\hat{\mathcal A}
 \end{cases}$ \\otherwise: $\begin{cases}
     \bPhi^{(j+1)} = \bPhi^{(j)}\\
     \mathcal A^{(j+1)}=\mathcal A^{(j)}
 \end{cases}$ };
\begin{scope}[on background layer]
  \node[groupRIS,fit=(dIn)(phase)(score)(select)(amp)(construct)(check)(accept),
        inner xsep=3mm,inner ysep=1mm] (Gproj) {};
\end{scope}
\node[captionnode] at ([yshift=1mm]Gproj.north)
 {Hard projection and candidate acceptance};
\draw[flow] (dIn.east) |- (phase.west);
\draw[flow] (dIn.east) |- (score.west);
\draw[flow] (score.east) -- (select.west);
\draw[flow] (select.east) -- (amp.west);
\draw[flow] (amp.north) -- (construct.south);
\draw[flow] (phase.east) -- (construct.west);
\draw[flow] (construct.east) -- (check.west);
\draw[flow] (check.east) -- (accept.west);
\draw[zoom] (proj.south west) -- (Gproj.north west);
\draw[zoom] (proj.south east) -- (Gproj.north east);

\end{tikzpicture}%
}
\endgroup
\caption{Structure of the deep unfolded AO framework.}
\label{fig:aodu_diagram}
\vspace{-0.3cm}
\end{figure*}

\blue{Algorithm~\ref{alg:ao_main} summarizes the overall procedure of the proposed method, while Fig.~\ref{fig:aodu_diagram} illustrates it.
Starting from an initial feasible solution, the algorithm alternately updates the power allocation and the RIS coefficients until convergence, based on Algorithms~\ref{alg:power_allo} and~\ref{alg:deep_unfolding_hr_ris}, respectively. Specifically, for fixed RIS coefficients, we first optimize the power allocation matrix. Then, for the resulting power allocation, the RIS coefficients and the active set are updated.
The acceptance conditions~\eqref{eq:inference_feasibility} ensure that each accepted RIS update does not decrease the approximate EE in~\eqref{eq_new_EE_approx} for fixed power allocation, while preserving the feasibility of the previous power allocation. Hence, the subsequent Dinkelbach-based power-allocation step cannot decrease the approximate EE either, provided that the subproblem is solved optimally. Consequently, the AO algorithm generates a non-decreasing and upper-bounded sequence of approximate EE values, which converges in objective value.}

For the RIS updates, the dominant cost in each deep unfolded layer arises from the gradient computation in \eqref{eq:gradientris} and the subsequent projection operations. The gradient computation has complexity $\mathcal{O}(KN^2+KNM+MK^2+K^3+NK^2)$. Since $N$ is typically much larger than both $K$ and $M$, the dominant term is $\mathcal{O}(KN^2)$. The projection steps, including phase projection, Top-$N_{\mathsf{a}}$ selection, and amplitude clipping, require at most $\mathcal{O}(N\log N)$ operations, which is negligible compared with $\mathcal{O}(KN^2)$ for large $N$. Hence, the complexity of Algorithm~\ref{alg:deep_unfolding_hr_ris} is approximated as
 $\mathcal{O}\!\big(JKN^2\big),$
where $J$ denotes the number of unfolded layers. 
Since the power allocation step has lower complexity in the considered large $N$, the overall complexity of Algorithm~\ref{alg:ao_main} is dominated by the RIS updates and can be approximated as
$\mathcal{O}\!\left(SJKN^2\right),$
where $S$ denotes the number of outer AO~iterations. 






\begin{algorithm}[t]\small
\caption{\small Deep Unfolded AO Framework for~\eqref{opt_ao}}
\label{alg:ao_main}
\begin{algorithmic}[1]
    \Require $\mHt$, $\mHr$

    \State Initialize $\mP^{(0)}$, $\bPhi^{(0)}$, $\mathcal{A}^{(0)}$ , $s = 0$
    \For{$s = 1, \ldots, S$} 
        \State Apply Algorithm \ref{alg:power_allo} to obtain $\mP^{(s)}$ 

        \State Apply Algorithm \ref{alg:deep_unfolding_hr_ris} to obtain $\bPhi^{(s)}$, $\mathcal{A}^{(s)}$
    \EndFor
    \State \Return $\mP^\star = \mP^{(S)}$, $\bPhi^\star = \bPhi^{(S)}$, and $\mathcal{A}^\star = \mathcal{A}^{(S)}$
\end{algorithmic}
\end{algorithm}

\section{Conventional PGA Method} \label{sec:grad}
In the previous section, we proposed a deep unfolded AO-based solution to the main problem \eqref{op_main_1}. To establish a rigorous benchmark, this section presents a baseline approach based on PGA, where the beamforming and RIS coefficient matrices are updated directly based on PGA to solve the original problem \eqref{op_main_1}.
Unlike the AO framework, PGA updates all variables simultaneously and therefore requires the constraints to be handled through explicit projection operators. Since constraint~\eqref{eq:maxeed} cannot be directly enforced through projection, we incorporate it into the objective function using a penalty formulation~\cite{nguyen2024joint}. Accordingly, problem~\eqref{op_main_1} is reformulated~as
\vspace{-5pt}
\begin{subequations}\label{op_main_pga}
    \begin{align} 
       \quad \underset{\substack{ \mF, \bPhi, \mathcal{A} }}{\textrm{maximize}} \quad &  \mathsf{EE} + \sum_{k=1}^K \omega_k \ln{(\mathsf{SE}_k - \mathsf{SE}^{\mathsf{min}}_k)} \label{eq:pga_a} \\
        \textrm{subject to} \quad
        &\eqref{eq:maxeeb}-\eqref{eq:maxeec},\eqref{eq:maxeee}-\eqref{eq:maxeeg},
\label{eq:pga_b} 
    \end{align}
\end{subequations}
where $\omega_k > 0, \forall k$ are the weights of penalty terms. 
To apply the PGA method, we first derive the gradients of the objective function in \eqref{op_main_pga} with respect to $\mF$ and $\bPhi$, whose closed-form expressions are provided in Lemma~\ref{lem:1}. \blue{Since the binary phase-shift and active/passive selection constraints are discrete and prevent direct gradient computation, we adopt the continuous relaxation described in Section~IV-B2. Specifically, we denote the relaxed RIS coefficient matrix by $\bar{\bPhi}$ and temporarily treat all RIS elements as active during gradient computation, i.e., $\mathcal{A}\equiv\mathcal{N}$. After each gradient update, the original hybrid-RIS structure is restored through hard projection.
\begin{lemma} \label{lem:1}
Under the above continuous relaxation, the gradients of $\mathsf{EE}$ with respect to $\mF$ and $\bar{\bPhi}$ are given in \eqref{eq:gradf} and \eqref{eq:gradgamma}, respectively. Similarly, the gradients of $\mathsf{SE}_k$ with respect to $\mF$ and $\bar{\bPhi}$ are provided in \eqref{eq:rkgradf} and \eqref{eq:rkgradgamma}, respectively.
\end{lemma}
}
\begin{align}
    &\nabla_{\!\mF}\mathsf{SE}_k \!=\! \frac{2}{\ln\!2}(\!\frac{\bar{\mH}_k\mF}{\tr{\mF \mFh \bar{\mH}_k}\!+\! \hat{\sigma}_k^2}\!\!-\!\!\frac{\bar{\mH}_k\mF_{\bar{k}}}{\tr{\mF_{\bar{k}}\mF_{\bar{k}}^{\mathsf{H}} \bar{\mH}_k}\!+\! \hat{\sigma}_k^2}\!),
    \label{eq:rkgradf}
    \\
    &\nabla_{\!\bar{\bPhi}}\mathsf{SE}_k \!=\!\frac{2}{\ln\!2}\!(\!\frac{\mT_k\bar{\bPhi} \mG}{\tr{\!\bar{\bPhi}^{\mathsf{H}} \mT_k\bar{\bPhi} \mG \!}\!+\! \sigma_k^2}\!\!-\!\!\frac{\mT_k\bar{\bPhi} \mG_k}{\tr{\!\bar{\bPhi}^{\mathsf{H}} \mT_k\bar{\bPhi} \mG_k\!}\!+\! \sigma_k^2}\!)\!\circ \!\mI_N\!.
    \label{eq:rkgradgamma}
\end{align}
\begin{align}
    &\nabla_{\!\mF} \mathsf{EE}\! = \!\frac{\sum_{k=1}^K \!\nabla_{\mF} \mathsf{SE}_k\cdot P_{\mathsf{tot}}-\mathsf{SE}\cdot\left(\frac{2}{\xi}\mF+\frac{2}{\zeta}\mHt^{\mathsf{H}}\bar{\bPhi}^{\mathsf{H}}\bar{\bPhi}\mHt\mF\right)}{(P_{\mathsf{tot}})^2}
    \label{eq:gradf}\\
    &\nabla_{\!\bar{\bPhi}} \mathsf{EE} \!= \!\dfrac{\sum_{k=1}^K \!\nabla_{\bar{\bPhi}} \mathsf{SE}_k\cdot P_{\mathsf{tot}}-\mathsf{SE}\cdot\left(\frac{2}{\zeta}\bar{\bPhi}\mG\circ \mI_N\right)}{(P_{\mathsf{tot}})^2}
    \label{eq:gradgamma}
    \end{align}
where
\begin{align}
    \bar{\mH}_k &\triangleq \mHt^{\mathsf{H}}\bar{\bPhi}^{\mathsf{H}} \vhrk \vhrkH \bar{\bPhi} \mHt,
    \quad \mT_k \triangleq \vhrk\vhrkH
    \\
    \tilde{\mH} &\triangleq \mHt\mF\mFh\mHt^{\mathsf{H}}
    ,
    \quad \quad \quad \quad \tilde{\mH}_k \triangleq \mHt\mF_{\bar{k}}\mF_{\bar{k}}^{\mathsf{H}}\mHt^{\mathsf{H}}
    \\
    \mG &\triangleq \tilde{\mH} + \sigma_{\mathsf{R}}^2\mI, \quad \mG_k \triangleq \tilde{\mH}_k + \sigma_{\mathsf{R}}^2\mI
    \\
    \hat{\sigma}_k^2 &\triangleq \sigma_k^2 + \sigma_{\mathsf{R}}^2 \norm{\vhrkH \bar{\bPhi}}^2 = \sigma_k^2 + \sigma_{\mathsf{R}}^2 \tr{\bar{\bPhi}^{\mathsf{H}} \vhrk\vhrkH\bar{\bPhi}}
\end{align}
with $\mF_{\bar{k}}$ obtained by replacing the $k$-th column of $\mF$ with all zeros.
See Appendix \ref{sec:app1}.

Using the above gradients, the corresponding PGA updates are given by

\begin{align}
    \hat{\bPhi}^{(t+1)} \!&\!= \bPhi^{(t)} \!\!+ \!\nu^{(t)}_{\bPhi} 
     \!(\nabla_{\bar{\bPhi}} \mathsf{EE}\! +\! \!\sum_{k=1}^K \!\frac{\omega_k \!\cdot\!\nabla_{\bar{\bPhi}}\mathsf{SE}_k}{\mathsf{SE}_k \!-\! \mathsf{SE}^{\mathsf{min}}_k})\! 
    \Big|^{\bar{\bPhi} = \bPhi^{(t)}}_{\mF = \mF^{(t)}},
    \label{eq:pgagamma}&&\\
    \bPhi^{(t+1)} &= \bPi_{\bPhi}(\hat{\bPhi}^{(t+1)}),
    \label{eq:policygamma}&&\\
    \hat{\mF}^{(t+1)} \!&\!=\! \mF^{(t)}\! \!+ \!\nu^{(t)}_{\mF} 
    \!(\nabla_{\mF} \mathsf{EE}\! +\!\! \sum_{k=1}^K\! \frac{\omega_k \!\cdot\!\nabla_{\mF}\mathsf{SE}_k}{\mathsf{SE}_k \!-\! \mathsf{SE}^{\mathsf{min}}_k}) 
    \!\Big|^{\bar{\bPhi} = \bPhi^{(t+1)}}_{\mF = \mF^{(t)}}\!,
    \label{eq:pgaf}&&\\
    \mF^{(t+1)} &= \bPi_{\mF}(\hat{\mF}^{(t+1)}),
    \label{eq:policyf}&&
\end{align}
where $\Pi_{\bPhi}$ and $\Pi_{\mF}$ denote projection operators that enforce the constraints of problem~\eqref{op_main_pga}. 
The projection $\Pi_{\bPhi}$ follows the same hard-projection procedure described in Section~\ref{subsec:deepunfold}. The projection $\Pi_{\mF}$ enforces the BS transmit-power constraint in~\eqref{eq:maxeeb} and the dynamic amplifier-power constraint in~\eqref{eq:maxeec}.


\section{Simulation Results} \label{sec:result}
In this section, we provide simulation results to demonstrate the proposed method.  We consider a scenario where the BS and RIS are located at $(0,0,10)\,\mathrm{m}$ and $(5,95,10)\,\mathrm{m}$, respectively. 
The $K=4$ users are distributed within a circular region of radius $10\,\mathrm{m}$ centered at $(0,100,0)\,\mathrm{m}$. This geometry represents a typical coverage-enhancement scenario in which the RIS assists users located in a clustered area with limited direct connectivity to the BS. The BS is equipped with a uniform linear array (ULA) of $M=8$ antennas arranged along the $x$-axis with half-wavelength spacing. 
The RIS employs a uniform planar array (UPA) with $N =N_y N_z = 10 \times 10$ elements arranged in the $y$–$z$ plane, also with half-wavelength~spacing.

The carrier frequency is set to $f_c = 3.5\,\mathrm{GHz}$. The noise power spectral density is $N_0 = -174\,\mathrm{dBm/Hz}$, and the system bandwidth is $20\,\mathrm{MHz}$, which results in a noise power of $\sigma_k^2 = \sigma_{\mathsf{R}}^2 = -101\,\mathrm{dBm}$. The BS circuit power is fixed to $P_{\mathsf{c,BS}} = 36.0\,\mathrm{dBm}$, and the power amplifier efficiency coefficients are set to $\xi = \zeta = 0.9$~\cite{ntougias2025hybrid}. Each RIS phase-shift controller consumes $P_{\mathsf{ps}} = -5.0\,\mathrm{dBm}$, while each active element requires an additional bias power of $P_{\mathsf{dc}} = -5.0\,\mathrm{dBm}$. The maximum amplitude coefficient of each active RIS element is set to $\alpha^{\mathsf{max}}=100$~\cite{long2021active}. The minimum SE requirement for each user is set to $\mathsf{SE}_k^{\mathsf{min}} = 0.1\,\mathrm{bits/s/Hz}$. Unless otherwise stated, we set the number of RIS elements to $N = 100$, with $N_{\mathsf{a}} = 60$ active elements, and the BS and RIS transmit power budgets to $P_{\mathsf{t}}^{\mathsf{max}} = 30\,\mathrm{dBm}$ and $P_{\mathsf{r}}^{\mathsf{max}} = -10\,\mathrm{dBm}$, respectively.

\blue{The BS--RIS and RIS--user links are modeled as quasi-static narrowband single-carrier Rician fading channels~\cite{nguyen2022hybrid}. We assume that perfect CSI is available for the considered channels, although acquiring CSI in practical RIS-assisted systems remains challenging. Specifically, the BS--RIS channel is given~by
\begin{equation}
    \mH_{\mathsf{t}}
    =
    \sqrt{\beta_{\mathsf{t}}(d_{\mathsf{t}})}
    \left(
    \sqrt{\frac{\kappa_{\mathsf{t}}}{1+\kappa_{\mathsf{t}}}}\,
    \mH_{\mathsf{t}}^{\mathsf{LoS}}
    +
    \sqrt{\frac{1}{1+\kappa_{\mathsf{t}}}}\,
    \mH_{\mathsf{t}}^{\mathsf{NLoS}}
    \right),
    \label{eq:htrician}
\end{equation}
where $\kappa_{\mathsf{t}}$ is the BS--RIS Rician factor, which is assumed to be $4$ in our simulations, and $\beta_{\mathsf{t}}(d)$ denotes the large-scale fading coefficient, given by $\beta_{\mathsf{t}}(d) = \beta_0\left(\frac{d}{d_0}\right)^{-\nu_{\mathsf{t}}}$, where $\beta_0$ denotes the free-space reference channel power gain, corresponding to $-43.33$ dB at $d_0=1$ m for the considered carrier frequency, and the BS--RIS path-loss exponent is set to $\nu_{\mathsf{t}}=2.2$~\cite{nguyen2022hybrid}. Moreover, $\mH_{\mathsf{t}}^{\mathsf{LoS}}$ represents the deterministic LoS component, while $\mH_{\mathsf{t}}^{\mathsf{NLoS}}$ contains i.i.d. entries distributed as $\mathcal{CN}(0,1)$. The LoS component is modeled as $\mH_{\mathsf{t}}^{\mathsf{LoS}} = \mathbf{a}_{\mathsf{RIS}}(\theta_{\mathsf{r}},\phi_{\mathsf{r}})\mathbf{a}_{\mathsf{BS}}^{\mathsf{H}}(\theta_{\mathsf{t}})$, where $\mathbf{a}_{\mathsf{BS}}(\theta_{\mathsf{t}})$ denotes the steering vector of the BS ULA with AoD $\theta_{\mathsf{t}}$, while $\mathbf{a}_{\mathsf{RIS}}(\theta_{\mathsf{r}},\phi_{\mathsf{r}})$ denotes the steering vector of the RIS UPA with azimuth and elevation AoAs $\theta_{\mathsf{r}}$ and $\phi_{\mathsf{r}}$, respectively. Similarly, the RIS--user channels follow the Rician fading model with $\kappa_{\mathsf r}=1$ and $\nu_{\mathsf{r}}=2.8$~\cite{nguyen2022hybrid}.
\\
The training, validation, and test channel realizations are generated independently according to the channel model described above. The model has $J=5$ unfolded layers and is trained over BS power budgets $P_t^{\mathsf{max}}\in[0,40]$ dBm, RIS power budgets $P_r^{\mathsf{max}}\in[-40,10]$ dBm, and numbers of active elements $N_{\mathsf a}\in[0,100]$, using the Adam optimizer with a learning rate of $0.01$ and a batch size of $100$, with $1000$ training and $200$ validation channel realizations over $1000$ epochs. Testing is performed over $500$ channel realizations. The initial step size is set to $\eta_{\bPhi}=0.5$. The trained model is applied to different system configurations without retraining.
}

\begin{figure}[!t]
    \centering
    \includegraphics[width=0.95\linewidth,height=5cm,keepaspectratio=false]{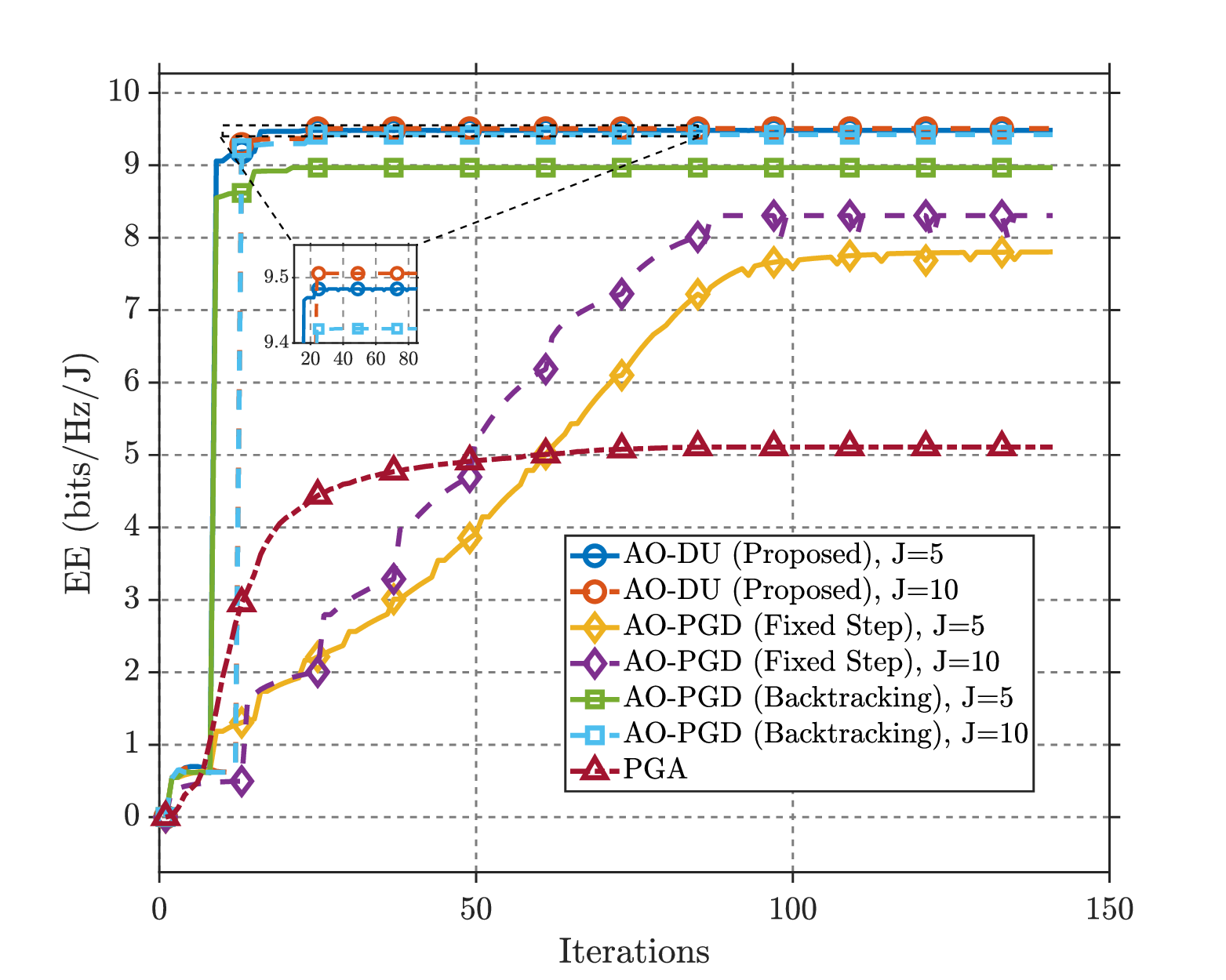}
    \vspace{-0.2cm}
    \caption{Convergence of AO-DU, AO-PGD, and PGA.}
    \label{fig:convergence}
\end{figure}

To assess the effectiveness of the proposed deep unfolding-based AO (AO-DU) method presented in Algorithm~\ref{alg:ao_main}, we compare it with the following benchmark schemes:
\begin{itemize}
    \item \textbf{AO-PGD}: Employs the same AO framework as the proposed method, but replaces the deep unfolding module with a conventional PGD algorithm to optimize the RIS coefficients. \blue{We consider both a fixed step size, denoted as \textbf{AO-PGD (Fixed Step)}, and a backtracking step-size selection, denoted as \textbf{AO-PGD (Backtracking)}.}

    \item \textbf{PGA}: Directly solves problem~\eqref{op_main_1} using the PGA method described in Section~\ref{sec:grad}.

    \item \textbf{Fixed}: Fixes the active element set $\mathcal{A}$ and optimizes the remaining variables using the proposed AO-DU scheme.

    \item \textbf{Random}: Randomly generates a feasible set of RIS variables once and keeps them fixed during the subsequent power allocation.

    \item \textbf{Fully-active}: Assumes that all RIS elements are active and applies the AO-DU algorithm to solve the resulting optimization problem.

    \item \textbf{Fully-passive}: Assumes that all RIS elements are passive and solves the problem using the AO-DU algorithm.
\end{itemize}
\blue{Fig.~\ref{fig:convergence} illustrates the convergence of the considered algorithms for solving problem~\eqref{op_main_1}. Each of the $S$ outer iterations consists of $I$ Dinkelbach iterations and $J$ RIS updates, resulting in $(I+J)S$ total iterations. We set $I=2$, $S=20$, and $J\in\{5,10\}$. As shown in Fig.~\ref{fig:convergence}, the proposed AO-DU converges rapidly and achieves the highest EE, followed by backtracking AO-PGD, while fixed-step AO-PGD converges more slowly and PGA achieves the lowest final EE. Table~\ref{tab:runtime} reports the achieved EE and the average online runtime under the same computational setup. Increasing $J$ from $5$ to $10$ provides only a marginal EE improvement for AO-DU, from $9.48$ to $9.51$ bits/Hz/J, while increasing its runtime from $0.66$ to $1.14$ s. Although backtracking AO-PGD approaches the EE of AO-DU for $J=10$, it requires a longer runtime. Therefore, we adopt $J=5$ in the remaining simulations.}
\begin{table}[!t]
\vspace{-0.4cm}
\caption{Achieved EE and Average Online Runtime of the Considered Algorithms with the Same Number of Iterations}
\label{tab:runtime}
\centering
\begingroup
\setlength{\tabcolsep}{3pt}
\renewcommand{\arraystretch}{1.0}
\begin{tabular}{lccc}
\hline
\textbf{Algorithm} & $\boldsymbol{J}$ & \textbf{EE (bits/Hz/J)} & \textbf{Runtime (s)} \\
\hline
AO-DU                     & 5  & 9.48 & 0.66 \\
AO-DU                     & 10 & 9.51 & 1.14 \\
AO-PGD (Fixed Step)       & 5  & 7.80 & 0.66 \\
AO-PGD (Fixed Step)       & 10 & 8.31 & 1.14 \\
AO-PGD (Backtracking)     & 5  & 8.97 & 0.94 \\
AO-PGD (Backtracking)     & 10 & 9.42 & 1.72 \\
PGA                       & -- & 5.11 & 1.71 \\
\hline
\end{tabular}
\endgroup
\vspace{-0.1cm}
\end{table}

\begin{figure}[!t]
    \vspace{-0.4cm}
    \centering
    \includegraphics[width=0.95\linewidth,height=5.5cm,keepaspectratio=false]{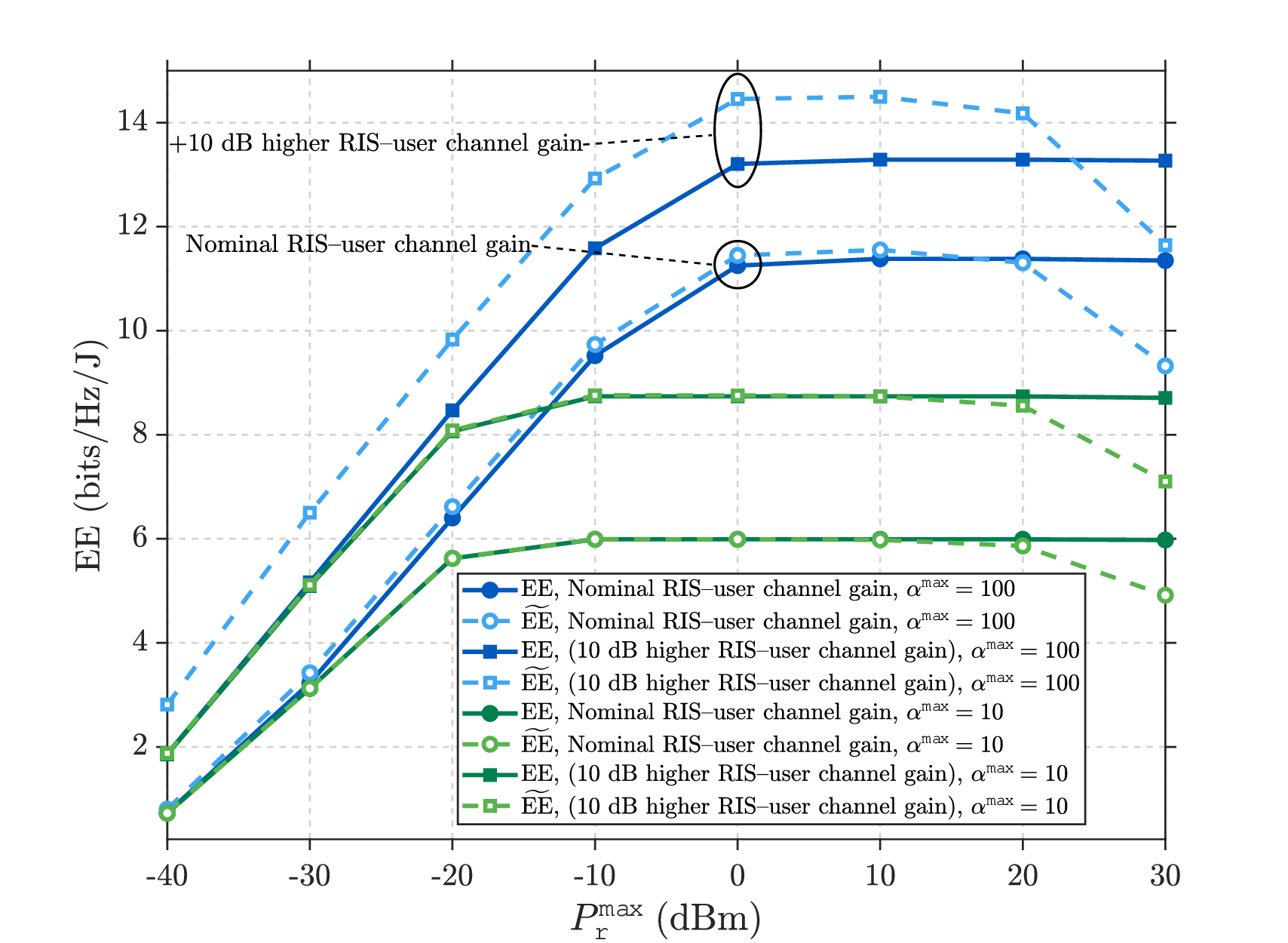}
    \vspace{-0.2cm}
    \caption{Exact EE and approximate $\widetilde{\mathsf{EE}}$ versus $P_{\mathsf{r}}^{\mathsf{max}}$ for the baseline and stronger RIS--user channel.}
    \label{fig:approx}
\end{figure}

\blue{
Fig.~\ref{fig:approx} compares the approximate $\widetilde{\mathsf{EE}}$ in \eqref{eq_new_EE_approx} with the exact EE in \eqref{eq_new_EE} for different RIS power budgets $P_{\mathsf r}^{\mathsf{max}}$. Both values are evaluated using the same solution obtained from Algorithm~\ref{alg:ao_main}, which optimizes the approximate objective. The approximation closely follows the exact EE, particularly at lower power budgets. To evaluate the impact of channel gain on the approximation accuracy, we also consider a 10 dB increase in the RIS--user channel gain. The higher channel gain increases both the achieved EE and the approximation gap due to stronger amplified RIS noise. Reducing $\alpha^{\mathsf{max}}$ from 100 to 10 limits amplification and reduces this gap. At higher RIS power budgets, the exact EE saturates, whereas the approximate EE decreases because it uses $P_{\mathsf r}^{\mathsf{max}}$ instead of the actual RIS power consumption, which is limited by the amplification constraints and the incident signal power. 
Overall, the results support the proposed approximation for practical low-power RIS operation. We also compare the proposed method with exhaustive search for a small-scale system and observe that it achieves EE close to the best solution found by exhaustive~search.}

\begin{figure}[t]
    \vspace{-0.1cm}
    \centering
    \includegraphics[width=0.9\linewidth,height=5.5cm,keepaspectratio=false]{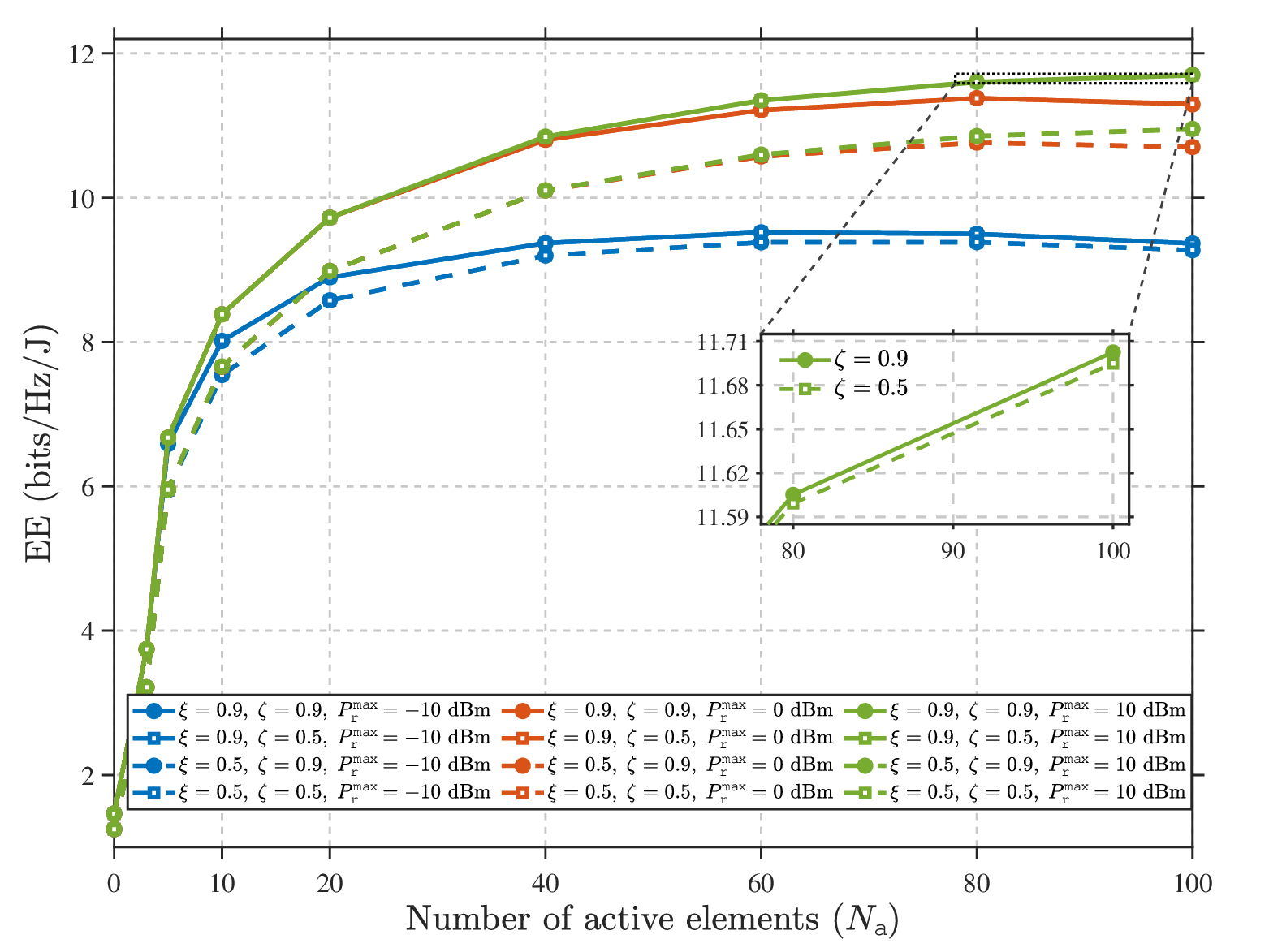}
    \vspace{-0.2cm}
    \caption{
    EE versus the predefined number of active elements $N_{\mathsf a}$ for $N=100$ under different RIS dynamic power budgets $P_{\mathsf r}^{\mathsf{max}}$.
    }
    \label{fig:ee_active_elements}
\end{figure}

\begin{figure}[!t]
    \vspace{-0.4cm}
    \centering
    \includegraphics[width=0.9\linewidth,height=5cm,keepaspectratio=false]{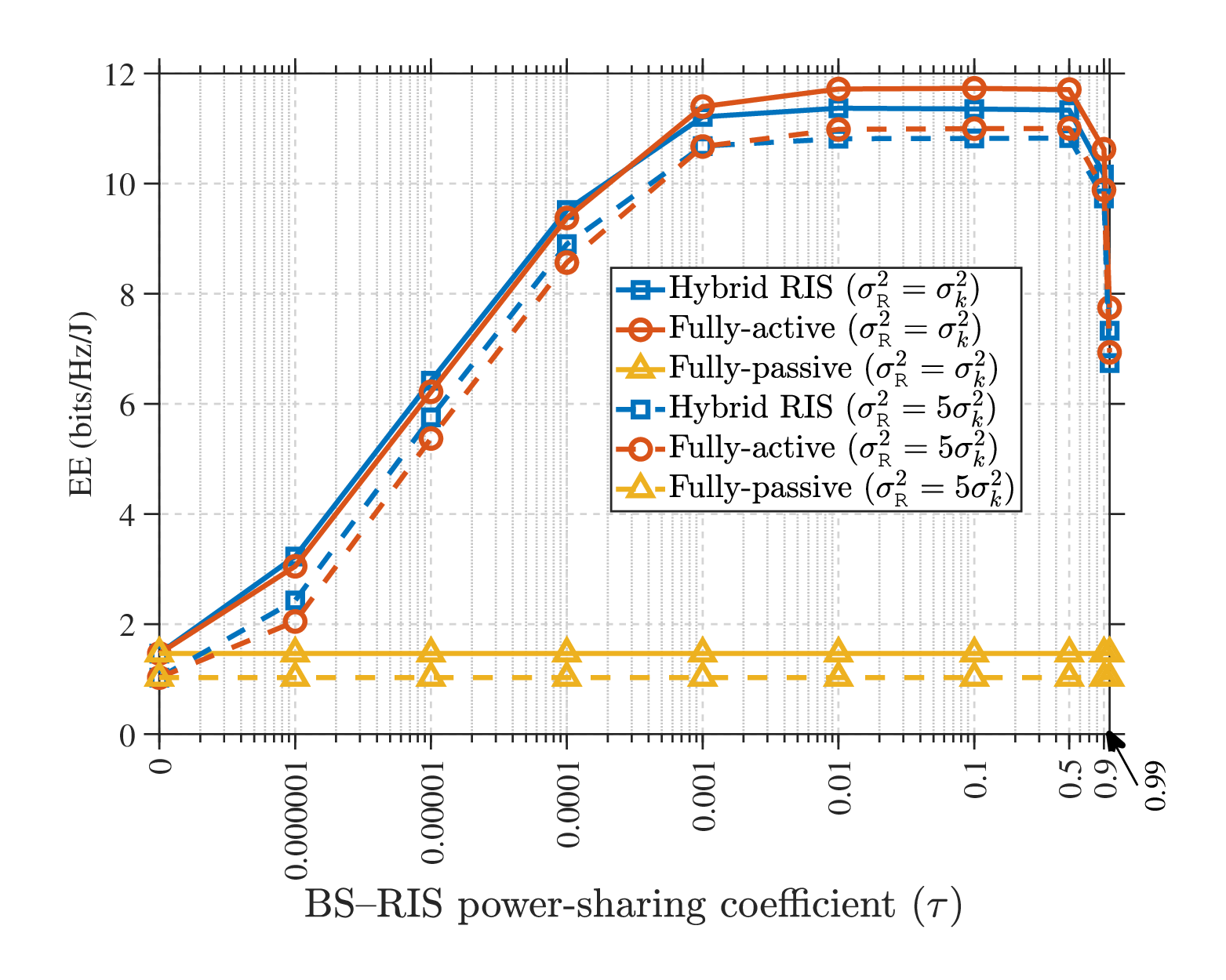}
    \vspace{-0.3cm}
    \caption{EE versus the power-sharing coefficient $\tau$ under a fixed total dynamic power budget (BS and RIS combined).}
    \label{fig:dynamic_power}
\end{figure}

\blue{Fig.~\ref{fig:ee_active_elements} shows the EE versus the predefined number of active elements $N_{\mathsf{a}}$ for different RIS dynamic power budgets $P_{\mathsf{r}}^{\mathsf{max}}$ and BS and RIS amplifier efficiencies $\xi,\zeta\in\{0.5,0.9\}$. For each fixed $N_{\mathsf{a}}$, the proposed method optimizes which RIS elements belong to the active set $\mathcal{A}$, subject to $|\mathcal{A}|=N_{\mathsf{a}}$. Here, $N_{\mathsf{a}}=0$ corresponds to the fully passive RIS, while $N_{\mathsf{a}}=100$ corresponds to the fully active RIS. For all considered power budgets and amplifier efficiencies, increasing $N_{\mathsf{a}}$ from zero improves the EE, since the active elements provide additional amplification gain and enhance the achievable rate. This improvement is significant even with a small number of active elements. For example, at $N_{\mathsf a}=20$, the EE reaches more than $80\%$ of its maximum value in all considered cases. However, as $N_{\mathsf a}$ increases further, the EE gain gradually saturates and may even decrease under limited RIS power budgets. This is because the available RIS dynamic power must be shared among more active elements, which limits their achievable amplitude coefficients. This effect is particularly evident for $P_{\mathsf{r}}^{\mathsf{max}}=-10$ and $0$ dBm, while for $P_{\mathsf{r}}^{\mathsf{max}}=10$ dBm, the EE continues to increase over a wider range of $N_{\mathsf{a}}$. Furthermore, reducing $\xi$ from $0.9$ to $0.5$ decreases the achieved EE due to higher BS power consumption. In contrast, reducing the RIS amplifier efficiency $\zeta$ from $0.9$ to $0.5$ has a negligible impact on EE.}
Fig.~\ref{fig:dynamic_power} illustrates the impact of the total dynamic power budget, defined as the sum of the BS transmit power budget and the RIS dynamic amplifier power budget, i.e.,
\begin{equation}
P_{\mathsf{tot}}^{\mathsf{dyn}} = P_{\mathsf{t}}^{\mathsf{max}} + P_{\mathsf{r}}^{\mathsf{max}}.
\end{equation}
This quantity represents the total available dynamic power that can be distributed between the BS and the RIS.
To examine the power-sharing effect, we proportionally allocate the total dynamic power between the BS and the RIS using a power-sharing coefficient $\tau$, such that
\begin{equation}
P_{\mathsf{r}}^{\mathsf{max}} = \tau P_{\mathsf{tot}}^{\mathsf{dyn}}, \qquad
P_{\mathsf{t}}^{\mathsf{max}} = (1-\tau) P_{\mathsf{tot}}^{\mathsf{dyn}},
\quad 0 \le \tau \le 1,
\label{eq:totdypower_from_r}
\end{equation}
where $\tau$ denotes the fraction of dynamic power allocated to the RIS.
\blue{We evaluate $\tau \in [0,0.99]$ for a fixed total dynamic power budget $P_{\mathsf{tot}}^{\mathsf{dyn}} = 30\,\mathrm{dBm}$ under three configurations: fully active RIS, fully passive RIS, and hybrid RIS.
To further evaluate the sensitivity of EE to RIS amplification noise, we also consider a fivefold increase in the RIS noise variance, i.e., $\sigma_{\mathsf{R}}^2=5\sigma_k^2$.
\\
In Fig.~\ref{fig:dynamic_power}, the EE of the fully passive RIS remains constant with respect to $\tau$, since it does not use dynamic amplification.
For the hybrid and fully active RIS architectures, increasing $\tau$ initially improves the EE.
However, with only a small portion of the total dynamic power, the hybrid RIS can already approach its maximum EE, since its limited number of active elements can effectively exploit the available amplification gain.
In contrast, the fully active RIS achieves a lower EE at very small values of $\tau$, but benefits more from increasing $\tau$ because the additional RIS power is distributed among a larger number of active elements.
Furthermore, increasing the RIS noise variance reduces the achievable EE of both architectures due to stronger amplified noise, while the overall power-sharing trends remain similar.
As $\tau$ increases further, the EE gain becomes limited and eventually decreases due to the power-sharing trade-off between the BS and the RIS.
In this regime, less power remains available at the BS, while the RIS cannot fully exploit the additional power due to the upper bound on the active-element amplitude coefficients.
Overall, the results show that allocating only a small fraction of the total dynamic power to the RIS is sufficient to approach the maximum EE, even under increased RIS amplifier noise.}
 

\begin{figure}[t!]
    \centering
    \vspace{-0.5cm}
    \subfloat[EE versus BS transmit power budget.]{
        \includegraphics[width=0.9\linewidth,height=5cm,keepaspectratio=false]{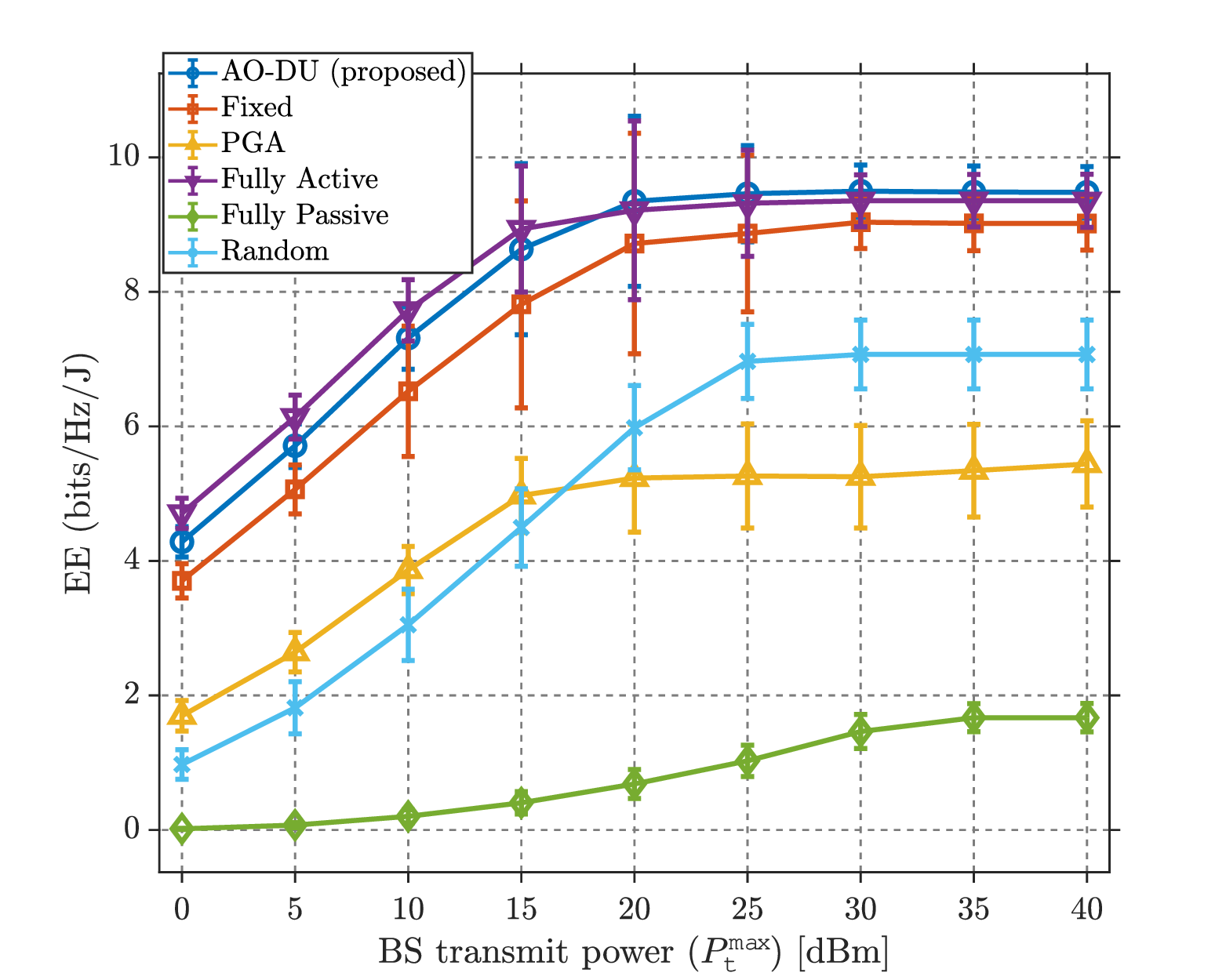}
        \label{fig:ee_vs_power}
    }
    \vspace{-0.3cm}
    \subfloat[SE versus BS transmit power budget.]{
        \includegraphics[width=0.9\linewidth,height=5cm,keepaspectratio=false]{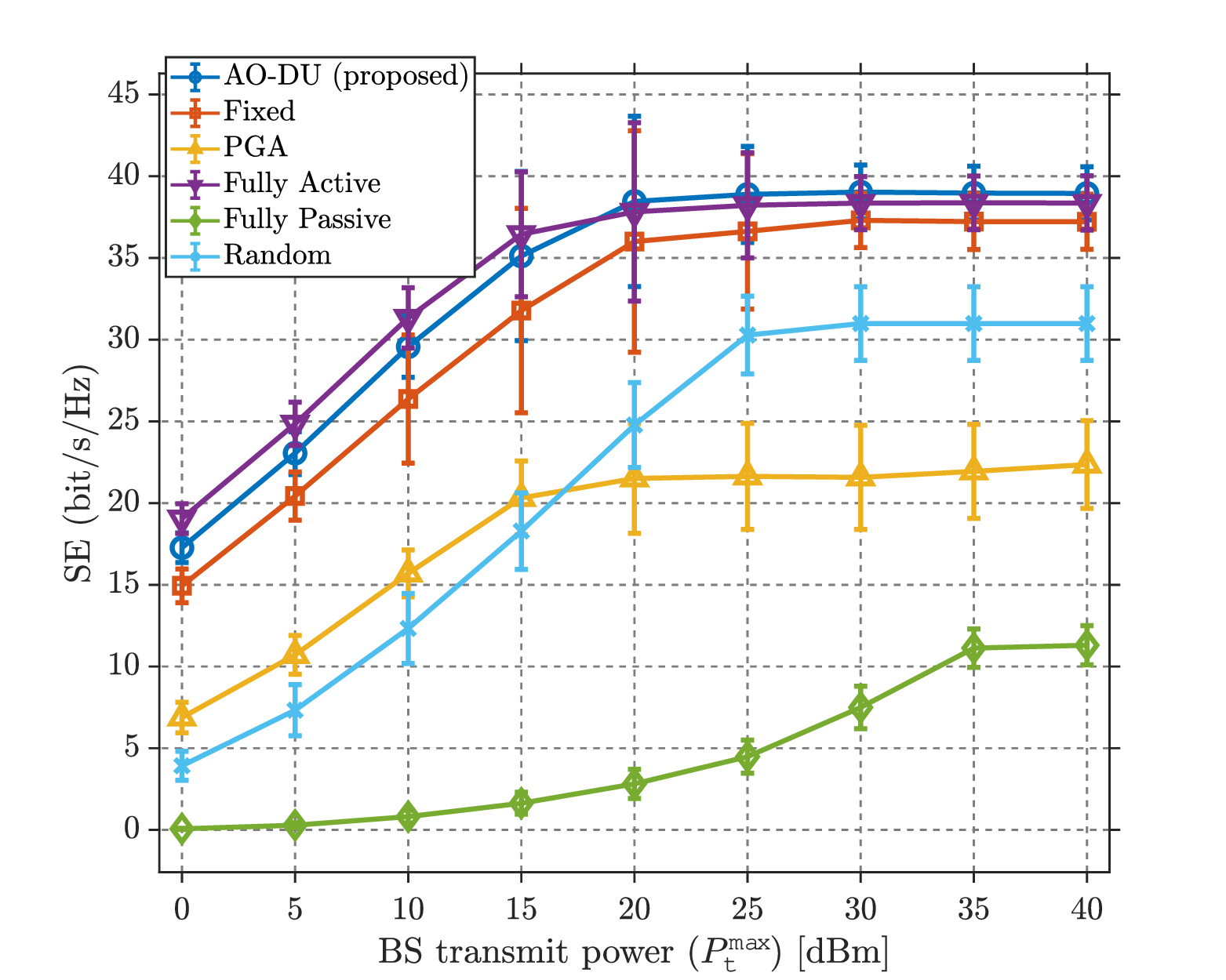}
        \label{fig:se_vs_power}
    }
    \caption{EE and SE versus the BS transmit power budget for different RIS configurations with $N=100$ elements.}
    \label{fig:ee_se_vs_power}
    \vspace{-0.1cm}
\end{figure}

\begin{figure}[!t]
    \centering
    \vspace{-0.4cm}
    \includegraphics[width=0.95\linewidth,height=5.5cm,keepaspectratio=false]{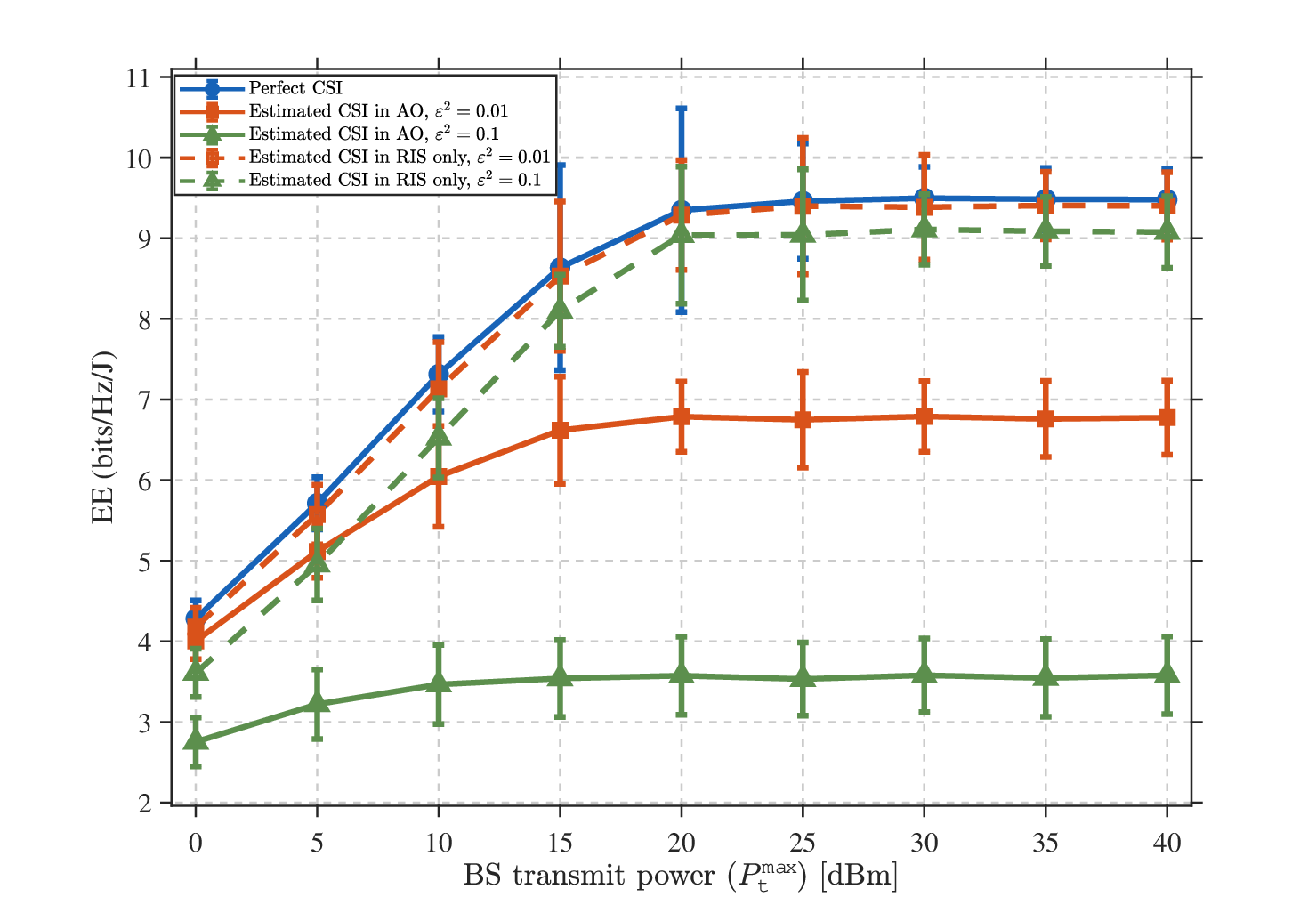}
    \vspace{-0.2cm}
    \caption{EE versus BS transmit-power budget under perfect and imperfect CSI for different channel estimation error levels.}
    \label{fig:imperfect_csi}
\end{figure}

\blue{
Figs.~\ref{fig:ee_vs_power} and~\ref{fig:se_vs_power} show the EE and SE versus the BS transmit power budget, respectively. The reported results are averaged over feasible channel realizations, with infeasible realizations excluded. All $500$ realizations are feasible except at $P_{\mathsf{t}}^{\mathsf{max}}=0$~dBm, where $358$ realizations satisfy the minimum SE constraint~\eqref{eq:maxeed}. Both EE and SE initially increase with $P_{\mathsf{t}}^{\mathsf{max}}$ and gradually saturate for the proposed AO-DU scheme, as EE maximization balances the achievable rate and power consumption without necessarily utilizing the entire available transmit power. It is also observed that the proposed AO-DU scheme achieves high performance in both figures over the considered transmit-power range. At low and moderate $P_{\mathsf{t}}^{\mathsf{max}}$, the Fully-Active RIS performs slightly better because the system is more transmit-power limited, and activating all RIS elements provides a stronger amplification gain and improves the effective cascaded channel. As $P_{\mathsf{t}}^{\mathsf{max}}$ increases, this advantage gradually diminishes because the BS power constraint becomes less restrictive and the EE-optimal solution does not necessarily use the entire available power budget. In this regime, the additional amplification gain of the Fully-Active RIS becomes less significant, while its higher amplifier power consumption and amplified noise remain. Consequently, the proposed hybrid AO-DU scheme becomes comparable to or better than the Fully-Active RIS at higher transmit-power budgets. The Fixed scheme performs worse than AO-DU, which highlights the importance of optimizing the active-element positions. The Random scheme also performs worse than these optimized schemes, while the Fully-Passive scheme has significantly lower performance, particularly in the high-power regime. The PGA benchmark also provides considerably lower performance than the proposed AO-DU scheme.
}
\\
\blue{Fig.~\ref{fig:imperfect_csi} illustrates the impact of imperfect CSI on the EE performance of the proposed AO-DU method versus the BS transmit-power budget. We consider additive channel estimation errors, given by ${\hat{\mathbf H}_{\mathsf t}=\mathbf H_{\mathsf t}+\Delta\mathbf H_{\mathsf t}}$ and ${\hat{\mathbf h}_{\mathsf r,k}=\mathbf h_{\mathsf r,k}+\Delta\mathbf h_{\mathsf r,k}}$, where $\mathbf H_{\mathsf t}$ and $\mathbf h_{\mathsf r,k}$ denote the true channels generated according to \eqref{eq:htrician}. The BS--RIS channel error is modeled as ${\Delta\mathbf H_{\mathsf t}=\sqrt{\beta(d_{\mathsf t})}\widetilde{\Delta\mathbf H}_{\mathsf t}}$, where the entries of ${\widetilde{\Delta\mathbf H}_{\mathsf t}}$ follow $\mathcal{CN}(0,\varepsilon^2)$ and $\beta(d_{\mathsf t})$ is the large-scale fading coefficient. The RIS--user channel errors are generated similarly, with $\varepsilon^2\in\{0.01,0.1\}$. Two cases are considered: (i) estimated CSI is used throughout AO and (ii) estimated CSI is used only in the RIS subproblem to assess the sensitivity of the deep-unfolding module, while perfect CSI is used for ZF beamforming, power allocation, and performance evaluation. Both cases include errors in the BS--RIS and RIS--user links.
\\
As shown in Fig.~\ref{fig:imperfect_csi}, imperfect CSI reduces EE, with greater degradation at higher error levels. At $P_{\mathsf t}^{\mathsf{max}}=30$ dBm, the EE decreases by approximately $28.5\%$ and $62.3\%$ for $\varepsilon^2=0.01$ and $0.1$, respectively, when estimated CSI is used throughout AO. In contrast, the corresponding reductions are only $1.2\%$ and $4.1\%$ when errors affect only the RIS subproblem. These results show that the deep-unfolding module maintains good EE performance under imperfect CSI without retraining, while the larger degradation in the overall AO procedure is mainly associated with CSI mismatch in ZF beamforming and power~allocation.}
\section{Conclusion} \label{sec:conc}
We addressed an EE maximization problem in a hybrid RIS-assisted MU-MISO downlink system under practical hardware constraints, including BS and RIS power budgets, active-element amplitude-coefficient limits, amplified RIS noise, and binary phase-shift control. The problem jointly optimizes the BS transmit beamforming, the positions and amplitude coefficients of a fixed number of active RIS elements, and the phase shifts of all RIS elements.
To solve this problem, we developed a deep-unfolded AO-based framework in which the BS beamforming subproblem is handled via ZF with closed-form power allocation, while the RIS subproblem is addressed using a model-driven deep-unfolding approach.

The numerical results showed that the proposed method achieves higher EE and faster convergence than the considered benchmark schemes. The results also confirmed that the hybrid RIS architecture achieves a more favorable rate-energy trade-off than the fully passive and fully active configurations and that only a limited number of active elements and a small portion of the dynamic power budget need to be assigned to the RIS to approach the best performance. These observations highlight the effectiveness of combining hybrid RIS design with deep unfolding for practical EE optimization. Future work may consider more general network settings, including multiple hybrid RISs, joint power sharing among the BS and multiple RISs, and robust system design under imperfect CSI.
\vspace{-0.3cm}
\appendices
\section{Proof of Lemma \ref{lem:1}}\label{sec:app1}
The following gradient derivation is carried out under the continuous
relaxation adopted in Section~\ref{sec:grad}, where $\mathcal{A}\equiv\mathcal{N}$ and $\bPhi_{\mathsf a}\equiv\bar{\bPhi}$ during the gradient calculation. We first compute the gradient of $\mathsf{EE}$ by applying the chain rule, which gives
\begin{align}
    \nabla_{\mF} \mathsf{EE} = \frac{\nabla_{\mF} \mathsf{SE} \cdot P_{\mathsf{tot}} - \mathsf{SE} \cdot \nabla_{\mF} P_{\mathsf{tot}}}{(P_{\mathsf{tot}})^2}
    \label{eq:chainf}
    \\ \nabla_{\bar{\bPhi}} \mathsf{EE} = \frac{\nabla_{\bar{\bPhi}} \mathsf{SE} \cdot P_{\mathsf{tot}} - \mathsf{SE} \cdot \nabla_{\bar{\bPhi}} P_{\mathsf{tot}}}{(P_{\mathsf{tot}})^2}
    \label{eq:chaingamma}
\end{align}
To derive the gradient of $\mathsf{SE}$, we first obtain the gradient of $\mathsf{SE}_k$ and then sum the resulting per-user gradients over all users. Based on \eqref{eq:spectral}, $\mathsf{SE}_k$ can be reformulated as
\begin{flalign}
    \mathsf{SE}_k &=  \log_2 \!\left( \frac{ \sum_{i=1}^{K}\abs{\vhrkH \bar{\bPhi} \mHt \vfi }^2+ \hat{\sigma}_k^2}{\sum_{i \neq k}^{K} \abs{ \vhrkH \bar{\bPhi} \mHt \vfi }^2 + \hat{\sigma}_k^2} \right) &&\nonumber\\&= \log_2 \!\left( \frac{ \tr{\mF \mFh\bar{\mH}_k}+ \hat{\sigma}_k^2}{\tr{\mF_{\bar{k}}\mF_{\bar{k}}^{\mathsf{H}}\bar{\mH}_k} + \hat{\sigma}_k^2} \right) 
    \label{eq:sef}&&
    \\&=\log_2 \left( \frac{ \tr{\bar{\bPhi}^{\mathsf{H}} \mT_k\bar{\bPhi} \mG }+ \sigma_k^2 }{\tr{\bar{\bPhi}^{\mathsf{H}} \mT_k\bar{\bPhi}\mG_k} + \sigma_k^2} \right),&&
    \label{eq:segamma}
\end{flalign}
where $\bar{\mH}_k$, $\tilde{\mH}$, and $ \tilde{\mH}_k$ are defined in Lemma \ref{lem:1}.

Using \eqref{eq:sef}, the gradient of $\mathsf{SE}_k$ with respect to $\mF$ is~\cite{hjorungnes2011complex}
\begin{align}
    \nabla_{\mF} \mathsf{SE}_k &=  \frac{\partial }{\partial \mF^*}\log_2\left(\tr{\mF \mFh\bar{\mH}_k}+ \hat{\sigma}_k^2 \right) \nonumber\\&- \frac{\partial }{\partial \mF^*}\log_2\left(\tr{\mF_{\bar{k}}\mF_{\bar{k}}^{\mathsf{H}}\bar{\mH}_k}+ \hat{\sigma}_k^2 \right)\nonumber\\&= 2\frac{\bar{\mH}_k\mF}{\ln2\left(\tr{\mF \mFh \bar{\mH}_k}+ \hat{\sigma}_k^2\right)}\nonumber\\&-2\frac{\bar{\mH}_k\mF_{\bar{k}}}{\ln2\left(\tr{\mF_{\bar{k}}\mF_{\bar{k}}^{\mathsf{H}} \bar{\mH}_k}+ \hat{\sigma}_k^2\right)}.
    \label{gradsekf}
\end{align}
Similarly, from \eqref{eq:segamma}, the gradient of $\mathsf{SE}_k$ with respect to $\bar{\bPhi}$ is
\begin{align}
    \nabla_{\bar{\bPhi}} \mathsf{SE}_k &=  \frac{\partial }{\partial \bar{\bPhi}^*}\log_2\left(\tr{\bar{\bPhi}^{\mathsf{H}} \mT_k\bar{\bPhi} \mG}+ \sigma_k^2\right) \nonumber\\&- \frac{\partial }{\partial \bar{\bPhi}^*}\log_2\left(\tr{\bar{\bPhi}^{\mathsf{H}} \mT_k\bar{\bPhi} \mG_k}+ \sigma_k^2\right)\nonumber\\&=2\frac{\mT_k\bar{\bPhi} \mG}{\ln2\left(\tr{\bar{\bPhi}^{\mathsf{H}} \mT_k\bar{\bPhi} \mG }+ \sigma_k^2\right)}\circ \mI_N\nonumber\\&-2\frac{\mT_k\bar{\bPhi} \mG_k}{\ln2\left(\tr{\bar{\bPhi}^{\mathsf{H}} \mT_k\bar{\bPhi} \mG_k}+ \sigma_k^2\right)}\circ \mI_N.
    \label{gradsekgamma}
\end{align}
Consequently, the gradients of $\mathsf{SE}$ with respect to $\mF$ and $\bar{\bPhi}$ are given, respectively, by
\begin{align}
    \nabla_{\mF} \mathsf{SE} = \sum_{k=1}^K \nabla_{\mF} \mathsf{SE}_k,
    \label{gradsef}
    \\
    \nabla_{\bar{\bPhi}} \mathsf{SE} = \sum_{k=1}^K \nabla_{\bar{\bPhi}} \mathsf{SE}_k.
    \label{gradsegamma}
\end{align}
The gradient of $P_{\mathsf{tot}}$, based on \eqref{eq:powertot}, is given by
\begin{align}
    \nabla_{\mF} P_{\mathsf{tot}} &=\frac{2}{\xi}\mF+\frac{2}{\zeta}\mHt^{\mathsf{H}}\bar{\bPhi}^{\mathsf{H}}\bar{\bPhi}\mHt\mF,
    \label{eq:ptotgradf}
    \\ \nabla_{\bar{\bPhi}} P_{\mathsf{tot}} &= \frac{2}{\zeta}\Big(\bar{\bPhi} (\tilde{\mH} + \sigma_{\mathsf{R}}^2\mI)\Big)\circ \mI_N= \frac{2}{\zeta}\bar{\bPhi} \mG\circ \mI_N.
    \label{eq:ptotgradgamma}
\end{align}
Finally, substituting \eqref{gradsef} and \eqref{eq:ptotgradf} into \eqref{eq:chainf} yields \eqref{eq:gradf}. Similarly, substituting \eqref{gradsegamma} and \eqref{eq:ptotgradgamma} into \eqref{eq:chaingamma} gives \eqref{eq:gradgamma}, which completes the proof of Lemma \ref{lem:1}.



\bibliographystyle{IEEEtran}
\bibliography{conf_short,IEEEabrv,Bibliography}

\end{document}

%% file: Definition.tex
\newcommand{\eqtext}[1]{\ensuremath{\stackrel{\text{#1}}{=}}}

\newcommand{\T}{{\scriptscriptstyle\mathsf{T}}}
\renewcommand{\H}{{\scriptscriptstyle\mathsf{H}}}

\newsavebox{\foobox}
\newcommand{\slantbox}[2][.3]
{%
	\mbox
	{%
		\sbox{\foobox}{#2}%
		\hskip\wd\foobox
		\pdfsave
		\pdfsetmatrix{1 0 #1 1}%
		\llap{\usebox{\foobox}}%
		\pdfrestore
	}%
}
\newcommand{\setA}{\mathcal{A}}

\definecolor{kugray5}{RGB}{224,224,224}

\newcommand\rsout{\bgroup\markoverwith
	{\textcolor{red}{\rule[0.5ex]{2pt}{0.8pt}}}\ULon}

\makeatletter
\newenvironment{breakablealgorithm}
{
		\begin{center}
			\refstepcounter{algorithm}
			\hrule height.8pt depth0pt \kern2pt
			\renewcommand{\caption}[2][\relax]{
				{\raggedright\textbf{\ALG@name~\thealgorithm} ##2\par}%
				\ifx\relax##1\relax 
				\addcontentsline{loa}{algorithm}{\protect\numberline{\thealgorithm}##2}%
				\else 
				\addcontentsline{loa}{algorithm}{\protect\numberline{\thealgorithm}##1}%
				\fi
				\kern2pt\hrule\kern2pt
			}
		}{
		\kern2pt\hrule\relax
	\end{center}
}
\makeatother

\newcommand{\blue}[1]{\textcolor{blue}{#1}}
\newcommand{\red}[1]{\textcolor{red}{#1}}
\newcommand{\green}[1]{\textcolor{green}{#1}}
\newcommand{\orange}[1]{\textcolor{orange}{#1}}

\newcommand{\myvec}[1]{\ensuremath{\begin{pmatrix}#1\end{pmatrix}}}

\makeatletter
\newcommand{\ALOOP}[1]{\ALC@it\algorithmicloop\ #1%
	\begin{ALC@loop}}
	\newcommand{\ENDALOOP}{\end{ALC@loop}\ALC@it\algorithmicendloop}
\newcommand{\algorithmicbreak}{\textbf{break}}
\newcommand{\BREAK}{\STATE \algorithmicbreak}
\makeatother

\let\mybibitem\bibitem
\renewcommand{\bibitem}[1]{%
	\ifstrequal{#1}{nature}
	{\color{blue}\mybibitem{#1}}
	{\color{black}\mybibitem{#1}}%
}

\graphicspath{ {Figures/} }

\newtheorem{definition}{\textbf{Definition}}
\newtheorem{theorem}{\textbf{Theorem}}
\newtheorem{remark}{\textbf{Remark}}
\newtheorem{lemma}{\textbf{Lemma}}
\newtheorem{corollary}{\textbf{Corollary}}
\newcommand{\dis}{\hspace{0.1cm}}

\renewcommand{\figurename}{Fig.}
\interdisplaylinepenalty=2500 

\newcommand\nbthis{\addtocounter{equation}{1}\tag{\theequation}}
\newcommand{\norm}[1]{\left\lVert#1\right\rVert} 
\newcommand{\normshort}[1]{\lVert#1\rVert} 
\newcommand{\eq}[1]{\begin{align*}#1\end{align*}} 
\newcommand{\eqn}[1]{\begin{align}#1\end{align}} 
\newcommand{\nt}[1]{\left(#1\right)} 
\newcommand{\nv}[1]{\left[#1\right]} 
\newcommand{\nn}[1]{\left\{#1\right\}} 
\newcommand{\abs}[1]{\left|#1\right|} 
\newcommand{\absshort}[1]{|#1|} 
\newcommand{\nb}{\numberthis}
\newcommand{\tr}[1]{\mathrm{tr}\left(#1\right)} 
\newcommand{\trshort}[1]{\mathrm{trace}(#1)} 

\newcommand{\detm}[1]{\mathsf{det}\left(#1\right)} 
\newcommand{\trsq}[1]{\mathsf{trace}^2\left(#1\right)} 
\newcommand{\diag}[1]{\mathrm{diag}\left\{#1\right\}} 
\newcommand{\var}[1]{\mathbb{V}\mathsf{ar}\left(#1\right)} 
\newcommand{\ds}{\displaystyle}
\newcommand{\re}[1]{\mathfrak{R}{\left(#1\right)}}
\newcommand{\im}[1]{\mathfrak{I}{\left(#1\right)}}
\allowdisplaybreaks
\newcommand{\cdf}{\mathbf{\textit{F}}} 
\newcommand{\diagtt}{\mathsf{diag}}
\newcommand{\pdf}{\mathbf{\textit{f}}} 
\newcommand{\mean}[1]{\mathbb{E} \left\{#1\right\}}
\newcommand{\meanshort}[1]{\mathbb{E} \{#1\}}

\newcommand{\cov}[1]{\mathbb{C} \left\{#1\right\}}
\newcommand{\meanBig}[1]{\mathbb{E} \Big\{#1\Big\}}
\newcommand{\covBig}[1]{\mathbb{C} \Big\{#1\Big\}}

\newcommand{\phybrid}{\left(\mathsf{P}_{\mathsf{hybrid}} \right)}
\newcommand{\pnhybrid}{\left(\mathsf{P-n}_{\mathsf{hybrid}} \right)}
\newcommand{\pnbhybrid}{\left(\var{\mathsf{Pn}}_{\mathsf{hybrid}} \right)}
\newcommand{\ppassive}{\left(\mathsf{P}_{\mathsf{passive}} \right)}
\newcommand{\pactive}{\left(\mathsf{P}_{\mathsf{active}} \right)}
\newcommand{\pnpassive}{\left(\mathsf{P-n}_{\mathsf{passive}} \right)}
\newcommand{\pnuniform}{\left(\mathsf{P-n}_{\mathsf{uniform}} \right)}
\newcommand{\pnhybridLB}{\left(\mathsf{P-n}_{\mathsf{fixed, uni.}} \right)}
\newcommand{\pdyn}{\left(\mathsf{P}_{\mathsf{dyn}} \right)}
\newcommand{\pdyna}{\left(\mathsf{P1}_{\mathsf{dyn}} \right)}
\newcommand{\prob}{\left(\mathsf{P} \right)}
\newcommand{\probn}{\left(\mathsf{Pn} \right)}

\newcommand{\mQ}{{\mathbf{Q}}}
\newcommand{\mR}{{\mathbf{R}}}
\newcommand{\mH}{{\mathbf{H}}} 
\newcommand{\mhsorted}{\underline{\mH}}
\newcommand{\mA}{{\mathbf{A}}}
\newcommand{\mS}{{\mathbf{S}}}
\newcommand{\mW}{{\mathbf{W}}}
\newcommand{\mP}{{\mathbf{P}}}
\newcommand{\mI}{\textbf{\textbf{I}}}
\newcommand{\mT}{{\mathbf{T}}}
\newcommand{\mB}{{\mathbf{B}}}
\newcommand{\mC}{{\mathbf{C}}}
\newcommand{\mD}{{\mathbf{D}}}
\newcommand{\mX}{{\mathbf{X}}}
\newcommand{\mY}{{\mathbf{Y}}}
\newcommand{\mG}{{\mathbf{G}}}
\newcommand{\mF}{{\mathbf{F}}}
\newcommand{\mU}{{\mathbf{U}}}
\newcommand{\mV}{{\mathbf{V}}}
\newcommand{\mE}{{\mathbf{E}}}
 \newcommand{\mHt}{{\mathbf{H}}_\mathsf{t}} 
\newcommand{\mIr}{{\mathbf{I}}_{N_r}}
\newcommand{\mZ}{{\mathbf{Z}}}
\newcommand{\mK}{{\mathbf{K}}}
\newcommand{\mN}{{\mathbf{N}}}
\newcommand{\mM}{{\mathbf{M}}}
\newcommand{\mRn}{{\mR_{\mathsf{n}}}}
\newcommand{\mRninv}{\mR_{\mathsf{n}}^{-1}}
\newcommand{\mJ}{{\mathbf{J}}}
\newcommand{\Na}{N_{\mathsf{A}}}
\def\mathbi#1{\textbf{\em #1}}

\newcommand{\Xm}{X_{max}} 
\newcommand{\Am}{A_{max}} 
\newcommand{\Bm}{B_{min}}

\newcommand{\setC}{\mathbb{C}} 
\newcommand{\setR}{\mathbb{R}}
\newcommand{\setD}{\mathcal{D}} 
\newcommand{\setANt}{\setA^{N_t}} 
\newcommand{\setAtld}{\tilde{\setA}^{N_t}}
\newcommand{\setN}{\mathbb{N}}

\newcommand{\ltabu}{\mathcal{L}}
\newcommand{\leta}{\mathcal{L}_{\eta}}
\newcommand{\lphi}{\mathcal{L}_{\phi}}

\newcommand{\Pe}{\mathcal{P}}
\newcommand{\Pne}{\var{\mathcal{P}}}

\newcommand{\vxb}{{\mathbf{x}}^{\star}}
\newcommand{\vc}{{\mathbf{c}}}
\newcommand{\vci}{{\mathbf{c}}_{\nn{1}}}
\newcommand{\ve}{{\mathbf{e}}} 
\newcommand{\vs}{{\mathbf{s}}}
\newcommand{\vx}{{\mathbf{x}}}
\newcommand{\vy}{{\mathbf{y}}}
\newcommand{\vr}{{\mathbf{r}}}
\newcommand{\vqn}{{\mathbf{q}}^T_n}
\newcommand{\vpnT}{{\mathbf{p}}^T_n}
\newcommand{\vpd}{{\mathbf{p}}_d}
\newcommand{\vrdc}{\check{{\mathbf{r}}}_d}
\newcommand{\vhdc}{\check{{\mathbf{h}}}_d}
\newcommand{\vhc}{\check{{\mathbf{h}}}}
\newcommand{\vrd}{{\mathbf{r}}_d}
\newcommand{\vrds}{{\mathbf{r}}_{d^{\star}}}
\newcommand{\vpn}{{\mathbf{p}}_n}
\newcommand{\vv}{{\mathbf{v}}}
\newcommand{\vn}{{\mathbf{n}}}
\newcommand{\vu}{{\mathbf{u}}}
\newcommand{\vz}{{\mathbf{z}}} 
\newcommand{\vh}{{\mathbf{h}}} 
\newcommand{\vhd}{{\mathbf{h}}_d}
\newcommand{\vq}{{\mathbf{q}}}
\newcommand{\vb}{{\mathbf{b}}}
\newcommand{\vw}{{\mathbf{w}}}
\newcommand{\va}{{\mathbf{a}}}
\newcommand{\vd}{{\mathbf{d}}}
\newcommand{\vt}{{\mathbf{t}}}
\newcommand{\vg}{{\mathbf{g}}}
\newcommand{\vp}{{\mathbf{p}}}
\newcommand{\vi}{{\mathbf{i}}}

\newcommand{\zbn}{\var{z}_n}
\newcommand{\rni}{r_{n,i}}
\newcommand{\rnj}{r_{n,j}}
\newcommand{\pnj}{p_{n,j}}
\newcommand{\qni}{q_{n,i}}
\newcommand{\qnj}{q_{n,j}}

\newcommand{\dx}{\Delta{\mathbf{x}}}

\newcommand{\mpeta}[1]{\phi_{#1} (\vx)} 
\newcommand{\mpetai}[1]{\phi_{#1} (\vx_{[i]})} 

\newcommand{\smt}{\sigma_t^2} 
\newcommand{\smv}{\sigma_v^2} 
\newcommand{\smn}{\sigma_n^2} 

\newcommand{\pbar}{\var{\Phi}}
\newcommand{\ptilde}{\tilde{\Phi}}

\newcommand{\bPhi}{\boldsymbol{\Phi}}
\newcommand{\bPi}{\boldsymbol{\Pi}}
\newcommand{\bUpsilon}{\boldsymbol{\Upsilon}}
\newcommand{\bTheta}{\boldsymbol{\Theta}}
\newcommand{\bPsi}{\boldsymbol{\Psi}}
\newcommand{\bSigma}{\boldsymbol{\Sigma}}
\newcommand{\bGamma}{\boldsymbol{\Gamma}}
\newcommand{\balpha}{\boldsymbol{\alpha}}
\newcommand{\Pa}{P_{\mathsf{a}}} 
\newcommand{\Pamax}{P_{\mathsf{a}}^{\mathsf{max}}}
\newcommand{\Pap}{P_{\mathsf{AP}}}
\newcommand{\amn}{\alpha_{mn}}
\newcommand{\amns}{\alpha_{mn}^{\star}}
\newcommand{\vphi}{\boldsymbol{\varphi}}
\newcommand{\vmu}{\boldsymbol{\mu}}
\newcommand{\vgamma}{\var{\boldsymbol{\gamma}}}
\newcommand{\veta}{\var{\boldsymbol{\eta}}}
\newcommand{\vf}{\mathbf{f}}
\newcommand{\vnu}{\boldsymbol{\nu}}
\newcommand{\vomega}{\boldsymbol{\omega}}
\newcommand{\valpha}{\boldsymbol{\alpha}}
\newcommand{\vtheta}{\boldsymbol{\theta}}

\def\redmark{\textcolor{red}}
\def\argmin{\mathop{\mathsf{argmin}}}
\def\argmax{\mathop{\mathsf{argmax}}}
\def\inf{\mathop{\mathsf{in}}}
\def\outf{\mathop{\mathsf{out}}}
\def\trace{\mathop{\mathsf{tr}}}
\def\dim{\mathop{\mathsf{dim}}}
\def\Re{\mathop{\mathsf{Re}}}
\def\Im{\mathop{\mathsf{Im}}}
\def\redmark{\textcolor{red}}

\newcommand{\Ei}{{\mathsf{Ei}}}

\def\bDelta{{\pmb{\Delta}}} \def\bdelta{{\pmb{\delta}}}
\def\bSigma{{\pmb{\Sigma}}} \def\bsigma{{\pmb{\sigma}}}
\def\bPhi{{\pmb{\Phi}}} \def\bphi{{\pmb{\phi}}}
\def\bGamma{{\pmb{\Gamma}}} \def\bgamma{{\pmb{\gamma}}}
\def\bOmega{{\pmb{\Omega}}} \def\bomega{{\pmb{\omega}}}
\def\bTheta{{\pmb{\Theta}}} \def\btheta{{\pmb{\theta}}}
\def\bepsilon{{\pmb{\epsilon}}} \def\bPsi{{\pmb{\Psi}}}
\def\obH{\overline{\bH}}\def\obw{\overline{\bw}} \def\obz{\overline{\bz}}\def\bmu{{\pmb{\mu}}}
\def\b0{{\pmb{0}}}\def\bLambda{{\pmb{\Lambda}}} \def\oc{\overline{\bc}}

\newcommand{\Frf}{\mF_\mathsf{RF}}
\newcommand{\Fopt}{\mF_{\mathsf{opt}}}

\newcommand{\Frft}{\tilde{\mF}_\mathsf{RF}}
\newcommand{\Fbb}{\mF_\mathsf{BB}}
\newcommand{\fbb}{\vf_\mathsf{BB}}

\newcommand{\Wrf}{\bW_\mathsf{RF}}
\newcommand{\Wbb}{\bW_\mathsf{BB}}
\newcommand{\Nrf}{N_{\mathsf{RF}}} 
\newcommand{\Nr}{N_\mathsf{r}}
\newcommand{\Nt}{N_\mathsf{t}} 
\newcommand{\Nu}{N_\mathsf{u}}
\newcommand{\Ntn}{N_{\mathsf{t},n}} 
\newcommand{\Ntnt}{\tilde{N}_{\mathsf{t},n}} 
\newcommand{\Ns}{N_\mathsf{s}}
\newcommand{\Nrh}{N_\mathsf{rh}}
\newcommand{\Nrv}{N_\mathsf{rv}}
\newcommand{\Nth}{N_\mathsf{th}}
\newcommand{\Ntv}{N_\mathsf{tv}}

\newcommand{\Nrhn}{N_{\mathsf{r},n}^\mathsf{h}}
\newcommand{\Nrvn}{N_{\mathsf{r},n}^\mathsf{v}}
\newcommand{\Nthn}{N_{\mathsf{t},n}^\mathsf{h}}
\newcommand{\Ntvn}{N_{\mathsf{t},n}^\mathsf{v}}
\newcommand{\Nms}{N_{\mathsf{MS}}}

\newcommand{\Nrfr}{N_{\mathsf{RF,r}}}
\newcommand{\Nrft}{N_{\mathsf{RF,t}}}
\newcommand{\Mnr}{M_{n,\mathsf{r}}}
\newcommand{\Mnt}{M_{n,\mathsf{t}}}
\newcommand{\ar}{\va_\mathsf{r}} 
\newcommand{\at}{\va_\mathsf{t}} 
\newcommand{\arh}{\va_\mathsf{r,h}} 
\newcommand{\arv}{\va_\mathsf{r,v}} 
\newcommand{\ah}{\va_\mathsf{h}} 
\newcommand{\av}{\va_\mathsf{v}}
\newcommand{\art}{\tilde{\va}_\mathsf{r}} 
\newcommand{\att}{\tilde{\va}_\mathsf{t}} 
\newcommand{\atht}{\tilde{\va}_\mathsf{t,h}} 
\newcommand{\atvt}{\tilde{\va}_\mathsf{t,v}}
\newcommand{\dat}{{\dot{\va}_\mathsf{t}}} 
\newcommand{\ih}{i_\mathsf{h}} 
\newcommand{\iv}{i_\mathsf{v}} 
\newcommand{\dath}{\dot{\va}_\mathsf{t,h}} 
\newcommand{\datv}{\dot{\va}_\mathsf{t,v}}

\newcommand{\Afull}{\mathcal{A}_{\mathsf{full}}}
\newcommand{\Asub}{\mathcal{A}_{\mathsf{sub}}}

\newcommand{\setK}{\mathcal{K}}
\newcommand{\setT}{\mathcal{T}}
\newcommand{\setJ}{\mathcal{J}}
\newcommand{\setM}{\mathcal{M}}
\newcommand{\setTheta}{\bm{\Theta_{\mathsf{d}}}}
\newcommand{\Pbs}{P_{\mathsf{BS}}}
\newcommand{\fronorm}[1]{\left\lVert#1\right\rVert_{\mathcal{F}}} 
\newcommand{\sigmac}{\sigma_{\mathsf{c}}^2}
\newcommand{\sigmas}{\sigma_{\mathsf{s}}^2}
\newcommand{\noiseu}{\sigma_{\mathsf{u}}^2}
\newcommand{\noiser}{\sigma_{\mathsf{r}}^2}
\newcommand{\noiserb}{\bar{\sigma}_{\mathsf{r}}^2}
\newcommand{\noise}{\sigma^2}
\newcommand{\Pt}{P_{\mathsf{t}}}

\newcommand{\Jtt}{J_{\theta \theta}}
\newcommand{\Jtp}{J_{\theta \phi}}
\newcommand{\Jta}{\mJ_{\theta \bar{\bm{\alpha}}}}
\newcommand{\Jpp}{J_{\phi \phi}}
\newcommand{\Jpa}{\mJ_{\phi \bar{\bm{\alpha}}}}
\newcommand{\Jaa}{\mJ_{\bar{\bm{\alpha}}\bar{\bm{\alpha}}}}
\newcommand{\Jpat}{\tilde{J}_{\phi \bar{\bm{\alpha}}}}

\newcommand{\xibf}{\bm{\xi}_{\mathsf{bf}}}
\newcommand{\lambdabf}{{\lambda_{\mathsf{bf}}}} 
\newcommand{\zetabf}{{\bm{\zeta}_{\mathsf{bf}}}}
\newcommand{\adotphi}{\dot{\va}_{\phi}}
\newcommand{\adottheta}{\dot{\va}_{\theta}}
\newcommand{\bdotphi}{\dot{\vb}_{\phi}}
\newcommand{\bdottheta}{\dot{\vb}_{\theta}}
\newcommand{\mHr}{{\mathbf{H}_{\mathsf{r}}}}
\newcommand{\mHrh}{{\mathbf{H}^{\mathsf{H}}_{\mathsf{r}}}}
\newcommand{\mHttidle}{{\tilde{\mathbf{H}}_{\mathsf{T}}}} 
\newcommand{\vhrk}{{\mathbf{h}}_{\mathsf{r}, k}}
\newcommand{\vhrkH}{{\mathbf{h}}_{\mathsf{r},k}^{\mathsf{H}}}
\newcommand{\vhrktilde}{{\tilde{\mathbf{h}}}_{\mathsf{R},k}}
\newcommand{\vhrktildeh}{{{\tilde{\mathbf{h}}}_{\mathsf{R},k}}^{\mathsf{H}}}
\newcommand{\mAT}{{\mathbf{A}^{\mathsf{T}}}}
\newcommand{\vfk}{\mathbf{f}_{k}}
\newcommand{\vfi}{\mathbf{f}_{i}}
\newcommand{\mFh}{{\mathbf{F}}^{\mathsf{H}}}
\newcommand{\bbeta}{\boldsymbol{\eta}}